\documentstyle[aps,graphicx]{revtex}

\begin{document} 

\title{{\LARGE\bf The century of the incomplete 
revolution: \\[1mm] 
searching for general relativistic quantum field theory}\\[3mm]
{\large \em Contribution to the  Special Issue 2000 of the 
Journal of  Mathematical Physics}\vskip.3cm} 
\author{{\bf Carlo Rovelli}\\
{\it Physics Department, University of Pittsburgh, 
Pittsburgh, PA 15260, USA;} \\ 
{\it Centre de Physique Th\'eorique, Case 907, Luminy, 
F-13288 Marseille, France.} 
\\[2mm]
rovelli@pitt.edu 
\vskip.2cm}
\date{\today}
\maketitle

\begin{abstract}
\noindent In fundamental physics, this has been the century of quantum
mechanics and general relativity.  It has also been the century of the
long search for a conceptual framework capable of embracing the
astonishing features of the world that have been revealed by these two
``first pieces of a conceptual revolution''.  I discuss the general
requirements on the mathematics and some specific developments towards
the construction of such a framework.  Examples of covariant
constructions of (simple) generally relativistic quantum field
theories have been obtained as topological quantum field theories, in
nonperturbative zero-dimensional string theory and its higher
dimensional generalizations, and as {\em spin foam models}.  A
canonical construction of a general relativistic quantum field theory
is provided by {\em loop quantum gravity}.  Remarkably, all these
diverse approaches have turn out to be related, suggesting an
intriguing general picture of general relativistic quantum physics.

\end{abstract} 

\section{New mathematics for fundamental physics}

\noindent In fundamental physics, the first part of the twentieth
century has been characterized by two important steps towards a major
conceptual revolution: quantum mechanics and general relativity.  Each
of these two theories has profoundly modified some key part of our
understanding of the physical world.  Quantum mechanics has changed
what we mean by matter and by causality and general relativity has
changed what we mean by ``where'' and ``when''.  The last part of the
century has then been characterized by the search for a new synthesis:
a unitary and comprehensive conceptual framework, capable of replacing
the Newtonian framework and embracing the astonishing features of the
world that have been revealed by quantum mechanics and by general
relativity.  Lacking a better expression, we can loosely denote a
theoretical framework capable of doing so as a ``background
independent theory'', or, more accurately, ``general relativistic
quantum field theory''.

The mathematics needed to construct such a theory must depart from the
one employed in general relativity --differentiable manifolds and
Riemannian geometry-- to describe classical spacetime, as well as from
the one employed in conventional quantum field theory --algebras of
local field operators, Fock spaces, Gaussian measures \ldots-- to
describe quantum fields.  Indeed, the first is incapable of accounting
for the quantum features of spacetime; the second is incapable of
dealing with the absence of a fixed background spatiotemporal
structure.  The new mathematics should be capable to describe the
quantum aspects of the geometry of spacetime.  For instance, it should
be able to describe physical phenomena such as the quantum
superposition of two distinct spacetime geometries, and it should
provide us with a physical understanding of quantum spacetime at the
Planck scale and of the ``foamy'' structure we strongly suspect it to
have.

Here, I wish to emphasize that what we have learned in this
century on the physical world --with quantum mechanics and
general relativity-- represents a rich body of knowledge which
strongly constraints the form of the general theory we are
searching.  If we disregard one or the other of these constraints
for too long, we just delay the confrontation with the hard
problems.  For example, string theory has developed to a large
extent disregarding the main physical lesson of general
relativity: as a result, string theorist are facing today the
problem of searching a genuinely background independent
formulation, thus realizing that after so many years of research
the structure of the fundamental theory is still unknown.

In the first part of this essay (Section \ref{sec2}), I give 
a general discussion of the main physical ``lessons'' we 
have learned from general relativity and from quantum field 
theory, and on the consequent constraints on the general 
form of the theory we are searching.  In particular, I 
discuss the physical meaning and the theoretical 
implications of diffeomorphism invariance, and I stress the 
fact that common concepts such as Poincar\'e invariance, or 
the hamiltonian, are weak field limit concepts that loose 
their physical significance when the gravitational field is 
strong.  They are therefore unlikely to play any role in the 
fundamental theory.

In the second part of this essay (Section \ref{sec3}), I
illustrate some of the ideas and the lines of research aimed at
developing a general relativistic quantum field theory.  A well
developed attempt to a rigorous construction of a general
relativistic quantum field theory is {\em loop quantum gravity},
a hamiltonian quantization of general relativity.  Loop quantum
gravity has obtained remarkable physical results on the Planck
scale quantum structure of space.  Of particular interest is the
derivation of the {\em discrete\/} spectrum of the physical area
a generic surface, and the physical volume of a generic spatial
region \cite{RovelliSmolin95}.  This result represents a set of
detailed and in principle falsifiable quantitative physical
predictions, it provides a physical interpretation for the spin
network states, the basis states of the theory
\cite{RovelliSmolin95b}, and gives us an intuitive picture of
quantum spacetime.

On the side of covariant formalisms, several examples of
generally covariant quantum field theories have been
constructed.  Some of these are very simple theories with a
finite number of degrees of freedom, such as the topological
quantum field theories \cite{Atiyah}.  Of particular
relevance here are combinatorial state sum constructions of
topological quantum field theories in three
\cite{Turaev-Viro} and four \cite{Turaev,Ooguri:1992b,Crane-Yetter}
dimensions.  Nonperturbative string theories ``in
zero-dimensions'', or 2d quantum gravity matrix models
\cite{2d} represent another simple example in low dimension. 
Matrix models have been generalized to three and four
dimensions, by Boulatov \cite{Boulatov:1992} and by Ooguri
\cite{Ooguri:1992b} in the form of 3d and 4d topological
quantum field theories.  There are several tentative
formulations of quantum GR itself as a model of this kind
\cite{Reisenberger97,Barrett:1998,othersn,ReisenbergerRovelli97},
some of which are based on the fact that GR can be seen as a
constrained form of $BF$ theory, a topological field theory. 
All these theories can be represented as {\em spin foam
models\/}: Feynman sums over 2-complexes (branched surfaces)
carrying spins \cite{Baez}.  Thus, spin foam models seem to
represent a generic covariant formalism for general
relativistic quantum field theory.

Remarkably, canonical and covariant approaches are related. 
A spacetime manifestly covariant formulation of loop quantum
gravity can be obtained by expanding the operator that
evolves the quantum states in the coordinate time, \`a la
Feynman.  This yields precisely a spin foam model.  The spin
foam represents in this case the history of the evolution of
a spin network.  Thus, a surprising number of very different
approaches converge towards a somewhat unitary description
of a general relativistic quantum field theory.  In this
formulation, a spin foam provides an intuitive picture of
the foamy features of the quantum spacetime geometry.

Here, I present only a brief view of some intriguing
developments and their connections.  For a more
comprehensive overview of current approaches to quantum
gravity, see \cite{Rovelli97b}.  My main aim is to show that
there is a field of converging ideas on the problem of
constructing general relativistic quantum field theory. 
Hopefully, these idea will lead us to a well defined
nontrivial theory whose classical limit is general
relativity; and thus to the conclusion of the beautiful
conceptual revolution opened at the beginning of the century
by quantum mechanics and general relativity.

\section{Why a general relativistic quantum field 
theory}\label{sec2}

\subsection{The lesson of general relativity: 
diffeomorphism invariance}\label{rel}

General relativity (GR) is the present theory of the
gravitational interaction.  It is a highly successful
theory, which in recent years has obtained spectacular
empirical support --binary pulsar's period decay due to
gravitational radiation, discovery of black holes in the
sky, \ldots--, has pervaded several branches of physics
--astrophysics, cosmology, \ldots--, and has even found
technological application --in the global positioning
system--, a development unthinkable not long ago.

However, GR is much more than just the theory of a specific 
physical force.  Indeed, GR is a theory of space and time.  
It has modified in depth our understanding of what space and 
time are, radically changing the Newtonian picture.  This 
modification of the basic physical picture of the world does 
not refer to the gravitational interaction alone.  Rather, 
it affects {\em any\/} physical theory.  Indeed, GR has 
taught us that the action of {\em all\/} physical systems 
must be generally covariant, not just the action of the 
gravitational field.  Thus, GR is a theory with a universal 
reach, whose implications involve the redefinition of our 
description of the whole of fundamental physics.

More in detail, GR has modified the physical meaning of the 
spacetime coordinates $x^{\mu}$ that enter our basic 
description of the world (as argument of the fields, or 
position of fundamental physical objects such as particles, 
strings, branes \ldots).  In Newtonian and special 
relativistic physics, the coordinates $x^{\mu}$ describe the 
spacetime localization of the events.  Events are thought to 
happen ``in'' spacetime.  Intuitively, spacetime can be 
thought as a stage over which physics happens.  Concretely, 
the localization of an event is determined with respect to a 
physical reference system, namely a set of physical objects 
chosen as spatiotemporal reference.  For instance, the value 
in $\vec x=0$ and $t=0$ of the electric field $\vec E(\vec 
x, t)$ represents a physical quantity that can be measured 
in a certain spacetime location determined by a physical 
reference system. 

If we take GR into account, this picture does not hold 
anymore, and the meaning of the coordinates $x^{\mu}$ is 
altered.  In fact, in a general relativistic theory physical 
quantities that have a coordinate dependence are not 
gauge-invariant.  Only quantities that {\em do not\/} depend 
on the coordinates may correspond to concretely physically 
observable quantities.  Localization with respect to a 
background spacetime, or with respect to a fixed external 
reference system, has no meaning.  What has physical meaning 
is only the {\em relative\/} localization of the dynamical 
objects of the theory (the gravitational field among them) 
with respect to one another.  The physical picture of the 
world provided by GR is not that of physical objects and 
fields over a spatiotemporal stage.  Rather, it is a more  
subtle picture of interacting entities (fields and 
particles) for which spatiotemporal coincidences only, and 
not spacetime localization, have physical significance. 
Once again, this modification of the meaning of the 
coordinates does not refer to the gravitational force alone: 
it refers to our entire description of the world at the 
fundamental level.

At the classical (non quantum) level, this novel view of 
space and time is expressed by the use of physical theories 
which are still defined over a ``spacetime'' differential 
manifold ${\cal M}$, but which are invariant under (active) 
diffeomorphisms $\phi:{\cal M}\to{\cal M}$ of the spacetime 
manifold ${\cal M}$ into itself.  The maps $\phi$ form a 
group, denoted $Diff_{\cal M}$.  More in detail, one first 
defines the fields $\varphi$ of the theory as if they were 
located over spacetime.  That is, as functions over the 
manifold $ \varphi:{\cal M}\to F, $ where $F$ is some 
field-value space.  Similarly, the dynamics of a particle is 
described by the worldline $X:R\to{\cal M}$ of the particle 
in ${\cal M}$.  Then, however, one chooses a diffeomorphism 
invariant action functional
\begin{equation}
\label{action}
      S[\varphi,X_{n}] = S[\phi(\varphi),\phi(X_{n})], 
           \hspace{2cm}  \forall\ \phi\in Diff_{\cal M},    
\end{equation}
where $\varphi$ represents all the fields and $X_{n},\ 
n=1\ldots N$ represents $N$ particles's worldlines.  In 
(\ref{action}), $Diff_{\cal M}$ acts geometrically on the 
space of the fields and of the particle trajectories.  (It 
acts on the dynamical variables of the theory only, not on 
fixed nondynamical structures.).  For instance, if $\varphi$ 
is a scalar field, $\phi(\varphi) = \varphi\circ\phi$, and, 
for the particle worldline, $\phi(X_{n}) = \phi\circ X_{n}$.

One of the main results of twentieth century's fundamental 
physics is that at the fundamental level the physical world 
is described by theories with property (\ref{action}). 

Property (\ref{action}) implies that spacetime localization 
is relational, for the following reason.  If 
$(\varphi,X_{n})$ is a solution of the equations of motion, 
then so is $(\phi(\varphi),\phi(X_{n}))$.  But $\phi$ might 
be the identity for all coordinate times $t$ before a given 
$t_{0}$ and differ from the identity for some $t>t_{0}$.  
The value of a field at a given point in $\cal M$, or the 
position of a particle in $\cal M$, change under the active 
diffeomorphism $\phi$.  If they were observable, determinism 
would be lost, because equal initial data could evolve in 
physically distinguishable ways respecting the equations of 
motion.  Therefore classical determinism forces us to 
interpret the invariance under $Diff_{\cal M}$ as a gauge 
invariance: we must assume that diffeomorphic configurations 
are physically indistinguishable.  That is
\begin{equation}
      (\varphi, X_{n}) \sim (\phi(\varphi),\phi(X_{n})),
\end{equation}
where $\sim$ means physically indistinguishable.  A 
(classical) physical gauge invariant state of the system $s$ 
is not described by a field configuration (or by the 
location of the particles), but rather by the equivalence 
class of field configurations (and particle locations), 
related by diffeomorphisms.

The quantities that have physical meaning, namely that can 
be predicted by the theory once the state is known, or 
whose measurement gives information on the state of the 
system, are diff-invariant quantities, that is, functions 
$Q$ of the dynamical variables $\varphi$ and $X$ that 
satisfy
\begin{equation}\label{diffinv}
    Q[\varphi,X_{n}]=Q[\phi(\varphi),\phi(X_{n})], 
           \hspace{2cm}  \forall\ \phi\,\in\,Diff_{\cal M}.  
\end{equation}
These quantities are the ``observables'' of a general
relativistic theory.  They do not have a dependence on the
coordinates $x^{\mu}$, and they are not functions on $\cal
M$.  Indeed, anything which depends on the coordinates
$x^{\mu}$ or is a functions on $\cal M$ is
gauge-noninvariant, and therefore does not represent a
physical quantity.  Examples of diff-invariant quantities
satisfying (\ref{diffinv}) are the Earth-Venus distance
during the last solar eclipses, the number of pulses of a
pulsar in a binary system that reach the Earth during one
revolution of the system (that is, between two Doppler
maxima), the energy deposited on a gravitational antenna by
a gravitational wave pulse and, in fact, any significative
physical quantity measured in general relativistic
experimental or observational physics.  For a more detailed
discussion on observability in GR, see 
\cite{Rovelli91b,Rovelli97c,Rovelli99}. 

I will say that an observable quantity in a field theory is
``weakly local'' if it is localized on the manifold, that
is, if it depends on the coordinates.  I will say that it is
``weakly nonlocal'' is it doesn't.  The adjective ``weak''
is to distinguish this notion of locality from others
employed in physics, such us spacelike commutativity of the
quantum fields or absence of nonlocal interactions. 
Observables in a general relativistic theory are weakly
nonlocal.  On the other hand, general relativistic
observables are typically ``local'' in a weaker, relational,
sense: for instance, the value of the Ricci scalar in the
spacetime point in which two particles' worldlines intersect
is local in the sense that it is localized in terms of
physical degrees of freedom, the particles.  I will say that
observables of this type are ``relationally local''. 
Typical GR observables are thus relationally local and
weakly nonlocal.

In a general relativistic context, the spacetime manifold 
${\cal M}$, whose points are labeled by the coordinates 
$x^{\mu}$, is thus nothing more than an auxiliary 
mathematical device for describing spatiotemporal relations 
between dynamical objects.  These relations are given by 
spatiotemporal coincidence, not by localization with respect 
to spacetime, or with respect to external reference system 
objects.  A displacement, or an arbitrary smooth deformation 
of all dynamical objects over the manifold $\cal M$ is a 
change of mathematical description, not a change of physical 
state.  Coordinate dependent quantities have no operational 
meaning.  Only quantities that do not depend on the 
coordinates have physical meaning.  Localization over $\cal 
M$ has no physical meaning.  Only localization of the 
dynamical objects with respect to one other has physical 
meaning.  This is a deep change in the way physics treats 
localization.  As we shall see below, the effects of this 
change on quantum field theory are considerable.

I conclude this section by briefly discussing a few 
important theoretical notions whose meaning has to be 
slightly generalized in order to make sense in a general 
relativistic context.  In particular, I will mention the 
phase space, the observable algebra and time.

The phase space $\Gamma$ is commonly defined as the space of 
the initial data at some initial time, up to gauge 
invariance, if any.  In general, however, $\Gamma$ admits a 
more covariant definition, as the space of the solutions of 
the equations of motion, up to gauge.  This definition makes 
sense in the general relativistic context.  Let $\cal C$ be 
the space of the solutions of the equations of motion of a 
diff invariant theory.  $Diff_{\cal M}$ acts on $\cal C$.  
The phase space of the theory is
\begin{equation}\label{quotient}
  \Gamma = \frac{\cal C}{Diff_{\cal M}}. 
\end{equation} 
$\Gamma$ carries a natural simplectic structure, determined 
by the action. Concretely, this simplectic structure can be 
computed  using the conventional hamiltonian framework.%
\footnote{In the hamiltonian framework, a solution of the 
equations of motion is coordinatized by its value on a 
(physically fictitious) ``initial value ADM hypersurface'', 
and the quotient (\ref{quotient}) is obtained by factoring 
away the gauge transformations generated by the Dirac, or 
ADM constraints $C^{\mu}$.  In this formalism,  
(\ref{diffinv}) becomes
\begin{equation}
	\{ Q,C^{\mu}\}=0.
	\label{ADM}
\end{equation}
An observable is a quantity having vanishing Poisson 
brackets with the ADM constraints.} %
However, it can also be directly computed in a covariant 
fashion%
.

The physical observables of the theory, such as the examples 
mentioned above, satisfy eq.(\ref{diffinv}), or 
eq.(\ref{ADM}), and are thus real functions on $\Gamma$.  An 
observable, indeed, has a well defined value on every 
physical state $s$.%
\footnote{Actually, observables of physical interest are 
often defined on {\em portions\/} of $\Gamma$ only, that is, 
only for certain classes of physical states.} %
In general, the space of the smooth%
\footnote{In general, $\Gamma$ fails to be a manifold, due 
in particular to the different dimensionality of the orbits 
of $Diff_{\cal M}$ in $\cal C$, and care must be accordingly 
taken in defining smoothness.  From the physical point of 
view, what matters are the not points of $\Gamma$, but open 
sets in $\Gamma$, because physical measurements have errors 
and the state is never exactly known.  Thus, in principle we 
can disregard the singular points of $\Gamma$ without 
changing physical predictions.  In practice these singular 
points --spaces with symmetries-- are often the only ones in 
which we are able to compute something!} %
functions on a phase space $\Gamma$, equipped with the 
Poisson brackets determined by $\Gamma$'s simplectic 
structure, is a noncommutative algebra ${\cal A}_{cl}$, the 
classical observables Poisson algebra.  A physical state 
determines (and can be identified with) a positive 
functional on the observable algebra
\begin{equation}
     s(q) = q(s),\ \ \hspace{1cm} s\in\Gamma, q\in{\cal 
     A}_{cl}.  
\end{equation} 
That is, a state can be viewed as an ensemble of values that 
the observables can take.  We can thus take the algebra of 
the observables ${\cal A}_{cl}$ and its positive 
functionals as the fundamental elements that define the 
theory.

In a non general relativistic theory, the dynamics is given
by assigning the hamiltonian or the action of of the
Poincar\'e group, on $\Gamma$.  In a general relativistic
theory there is, in general, no hamiltonian defined on
$\Gamma$.  Coordinate time evolution is a gauge, and is
washed away by (\ref{diffinv}).  The algebra of the
diffeomorphism invariant observables ${\cal A}_{cl}$ codes
the full dynamical content of the theory.  Of course, ${\cal
A}_{cl}$ is highly non-trivial.  Evolution in clock time is
described by the diff-invariant (and thus coordinate-time
independent) correlations between a physical variable $q$
and a clock variable $t$.  For each real $t$, $q(t)$ is a
diff invariant function on $\Gamma$ (See
\cite{Rovelli91b,time}).

On the other hand, the physical ``flowing'' of time is not
described by the theory.  In my opinion, such a flow is
thermodynamical in origin and is state dependent; this
``thermal time'' can be identified as the state dependent
flow defined by the (generic) statistical state $s$ over the
observables algebra
\begin{equation} \label{tomc}
     \frac{dq}{dt} = - \{q,\log{s}\}. 
\end{equation} 
In words, it is not the hamiltonian $H$ that defines a
thermal Gibbs state $s=\exp\{-\beta H\}$, but rather the
other way around: the statistical state $s$ in which the
system happens to be determines a ``hamiltonian'' $H\sim - 
\log{s}$, and therefore a time flow.  This point of view
allows one to develop a statistical mechanics of the
gravitational degrees of freedom, a problem which is still
open.  The idea is discussed and developed in
\cite{termo,alain}.

\subsection{The lesson of quantum mechanics and quantum 
field theory: weakly local operator algebras}\label{quantum}

I begin this section with some general considerations on 
quantum mechanics (QM).  This century has replaced Newtonian 
mechanics with QM as the fundamental mechanical theory.  QM 
is a curious theory, which we probably haven't fully 
understood yet.  The meaning of quantum mechanics is 
completely clear as far as the theory is applied to describe 
a quantum systems $S$ interacting with a ``classical'' or 
``macroscopic'' systems $O$, the ``measuring apparatus'', or 
``observing system''.  On the other hand, the physical 
meaning of quantum mechanics as a general mechanical theory 
of {\em all\/} dynamical systems is viewed by many as 
controversial.

I myself understand quantum mechanics as a theory that
describes the interaction between {\em any two \/} physical
systems.  Given two systems, $S$ and $O$, the way $S$
affects $O$ in the course of an interaction is described by
the quantum theory of $S$, where $O$ is formally regarded as
the classical measuring device, {\em whatever its physical
properties}.  This implies that the properties that $S$
manifests in interacting with $O$ are not necessarily the
same it manifests in interacting with another physical
system, say $O'$.  The last, in fact, may be affected by the
quantum properties of $O$ and in particular by the $O-S$
quantum correlations.  These do not affect the way $S$
interacts with $O$ alone and are not taken into account in
the quantum theory of $S$ alone.  It follows that all
(contingent) properties of a system are relative to another
system.  There are no absolute properties, or an absolute
state, of a system.  A statement such as ``the $z$ component
of the spin of the electron $S$ is up'' should be
interpreted as ``the $z$ component of the spin of the
electron $S$, {\em with respect to the physical system
$O$\/}, is up'' I have elaborated relational this point of
view on quantum mechanics in \cite{qm}, to which I refer for
more details and related references.

It has been suggested by some that this issue --the 
interpretation of quantum mechanics-- must be solved 
together with the problem of combing QM and GR. I do not 
think that this is the case.  In fact, QM is uncontroversial 
as long as a macroscopic classical physical system $O$, 
which could be used as measuring apparatus or observing 
system, is available.  And a system of this sort is 
certainly available (say, the Earth).  GR forbids us from 
using external physical reference systems, but the external 
reference system that serves to localize objects should not 
be confused with the apparatus that performs a quantum 
measurement.  A measuring apparatus can perform a quantum 
measurement of a diffeomorphism invariant quantity.

On the other hand, it seems to me that there could be some
deep connection between the relational aspect of the world
revealed by GR (localization is relative to other dynamical
objects) and the relational aspects that, I think,
characterize QM (states and outcome of measurements are
relative to the observing, or interacting, object).  After
all, an observing object is somewhere in spacetime and, as
we know from special relativity, any interaction requires
spatiotemporal coincidence.  The other way around,
spatiotemporal coincidence can only be revealed by means of
a physical, and thus quantum, interaction.  Therefore the
weaving itself of spacetime seems to be woven with a thread
of quantum interactions.  However, I think we are still too
far from a theory capable of describing the world so
intimately.  I think that it is more productive, today, to
remain grounded on what we know well about the world, that
is GR and QM, and simply search for a general relativistic
quantum theory.  Let me thus close this parentheses of
general considerations, and get back to what we have learned
from QM.

The main lesson of quantum mechanics is that the states of a
physical system have a (projective) complex linear
structure, such that, given any set of states $s_{i}$, in
which the measurement of a quantity $Q$ yields the results
$q_{i}$, there exist also states $ s = \sum_{i} \alpha_{i}
s_{i} $ , with $ \sum_{i} |\alpha_{i}|^{2} = 1 $, in which
the quantity $Q$ can take any value $q_{i}$, each with
probability $|\alpha_{i}|^{2}$.  The linear space $H$ of the
states has a Hilbert structure and, for every observable
quantity $Q$, there is an orthonormal basis $s_{n}$ in which
$Q$ is determined and has value $q_{n}$.  Thus, for any $Q$
a self-adjoint operator $\hat Q$ on $H$ is defined by $\hat
Q s_{n}=q_{n}s_{n}$.  Since distinct observables in general
do not share eigenbases, the operators do not commute.  In
general, the set of the observables that characterize a
system forms a noncommutative operator algebra $\cal A$, the
quantum observable algebra, and $H$ provides a
representation of $\cal A$.  The algebra $\cal A$ is related
to the algebra ${\cal A}_{cl}$ of the classical theory that
describes the $\hbar\to 0$ limit.  In particular, $\cal A$
is a linear representation of (possibly a subalgebra of)
${\cal A}_{cl}$, or of a deformation (in $\hbar$) of the
same.  Thus Poisson brackets are remnants of quantum
noncommutativity.

This is the general form of a quantum mechanical system.  In the
course of the century, however, it has become increasingly clear that
nature can be described in terms of fields at the fundamental level. 
All the forces we know, as well as the special relativistic quantum
behavior of the elementary particles are well described by field
theories.  The present picture of nature, which is extraordinary
empirically successful, is thus a theory of fields.  The observable
quantities of a non general relativistic field theory are weakly
local: they are values of fields and local functions of these.  They
depend on the value of the fields at a point (or Fourier transforms of
the same), or in an open region of spacetime $\omega\subset{\cal M}$. 
A non general relativistic quantum theory of fields is a
representation of an algebra $\cal A$ of spacetime dependent
observables, or a {\em weakly local operator algebra}.  The set of the
observables with support on a region $\omega$ form a subalgebra ${\cal
A}_{\omega}$ and
\begin{equation}\label{locala} 
       \omega\subset\omega' \ 
       \Longrightarrow\ \  
       {\cal A}_{\omega}\subset{\cal A}_{\omega'}, 
       \hspace{1cm} \forall \ \ \omega,\omega'\subset{\cal M}, \ \  
       {\cal A}_{\omega},{\cal A}_{\omega'}\subset{\cal A}, 
\end{equation}
as subalgebras.  In particular, $\cal A$, with suitable
properties, may be an algebra of the quantum field operators
\cite{Haag}.  We denote a theory having this structure a
{\em weakly local quantum field theory}.

The quantum field theories that have proven so enormously 
successful in particle physics are weakly local quantum field 
theories.  Examples of observable in these theories are 
Whightman functions, scattering amplitudes, or the energy in 
a spatial region.  These quantities are defined on the 
spacetime manifold.  Physically, they describe a system 
which is {\em located\/} somewhere in spacetime, and which is 
studied by means of an external reference system that 
determines ``where'' and ``when'' measurements are 
performed.  The spacetime, or momentum, dependence of the 
quantum fields represents the spatiotemporal location of the 
field excitation.  More precisely, the spacetime dependence 
of the observables represents the spatiotemporal location of 
the apparata that reveal these field excitations.

String theory is a weakly local field theory of this type as
well.  The physical interpretation of the scalar fields
$X^{\mu}(\sigma,\tau)$ defined on the two-dimensional string
world sheet is the {\em location\/} of the string on the
spacetime manifold, the target space.  Thus, in order to
interpret the theory, we must assume that physical reference
systems exist in the target space.  The theory describes the
motion of the string with respect to these objects.  The
target space equipped with its fixed metric is a pre-general
relativistic ``absolute'' spacetime.%
\footnote{Clearly, the fact that the quantities 
$X^{\mu}(\sigma,\tau)$ are quantum operators does not mean 
that ``spacetime is quantized'', anymore than the fact that 
the position of a conventional particle in Minkowski space 
$X^{\mu}(\tau)$ is an operator does.} 

In conclusion, a quantum theory is defined by an operator 
algebra $\cal A$.  In a non general relativistic quantum 
{\em field\/} theory, $\cal A$ is a weakly local operator 
algebra.  Observables and states have a well defined 
spacetime dependence, which represents the spatiotemporal 
localization of the field excitations and of the measurement 
apparata.

\subsection{General relativistic quantum field theory}

We can now compare the two previous sections.  As discussed
in \ref{rel}, if we take relativistic gravity and GR into
account, we have to weaken the notion of localization.  The
observable quantities of a general relativistic theory are
weakly nonlocal.  Therefore they cannot form a weakly local
operator algebra of the kind (\ref{locala}).  Thus, the
mathematical structure of quantum field theory recalled in
the previous section is incompatible with the general
relativistic notion of localization described in Section
\ref{rel}.

A general relativistic quantum theory is a quantum field 
theory in which the observable algebra $\cal A$ and the 
states have no remnant of localization in spacetime.  Such 
a structure is very different from that of non general 
relativistic quantum field theory.  A quantum state, in 
particular, will not represent a field excitation 
``somewhere'' in space.  Localization has to be defined 
internally, with respect to the quantum states themselves.

The theory should then reduce to a conventional weakly local 
quantum field theory in the limit in which we disregard the 
quantum and dynamical properties of the gravitational field.  
In this limit, relational localization with respect to the 
gravitational field is reinterpreted as absolute 
localization, defined with respect to a fixed nondynamical 
spacetime.  In simpler words, if we do not consider the 
gravitational field as a dynamical entity, then we can use 
{\em it\/} as the background ``stage'' over which, and with 
respect to which, the other fields and the particles are 
localized, and move.

Is conventional QM sufficiently general to deal with
diffeomorphism invariance and with relational locality?  The
answer is subtle, and depends on what one precisely means by
conventional QM. In the Schr\"odinger picture, the states
$\psi(t)$ depend on time, and time evolution is governed by
the Schr\"odinger equation.  This structure is not general
enough to deal with general relativistic theories, in which
there is no external time-parameter evolution.  Indeed, an
external clock is precisely a minimal form of external
reference system.  However, QM can be formulated in the
Heisenberg picture as well.  In the Heisenberg picture, a
quantum state $\psi$ has no temporal connotations.  A
Heisenberg state is often viewed as the value of the
Schr\"odinger state at some fixed moment of time, but a more
useful view of a Heisenberg state is to see it as
representing the entire ``history'' of the system (at least,
until the next measurement).  Thus, a Heisenberg state is
the quantum analog of the classical state $s$ defined in
section \ref{rel} as a solutions of the equations of motion
up to gauge, without any reference to time.  In the
Heisenberg picture of a non-general relativistic theory,
observables have a time dependence, governed by a
hamiltonian $\hat H$
\begin{equation}\label{hei}
     \frac{d\hat Q}{dt} = i\hbar[\hat Q,\hat H].
\end{equation}
In a general relativistic theory there is no hamiltonian.  
Observables, however, must be diff-invariant, namely satisfy 
the quantum equivalent of equation (\ref{diffinv}).  In the 
hamiltonian framework, this means that the observable must 
satisfy the quantum version of (\ref{ADM}), that is
\begin{equation}\label{qADM}
	[\hat Q, \hat C^{\mu}]=0,
\end{equation}
where $ \hat C^{\mu}$ are the quantum constraints.  
Equation (\ref{qADM}) is the general relativistic 
generalization of equation (\ref{hei}).%
\footnote{Any hamiltonian theory admits a formulation in 
this more general framework.  For instance, the mechanics of 
a free particle in one dimension can be formulated over the 
four dimensional phase space $(X,P_{X},T,P_{T})$ with the 
single constraint $C = P_{T}+\frac{P^{2}}{2m}$, and no 
hamiltonian.  It is easy to see in this case that 
(\ref{qADM}) reduces precisely to (\ref{hei}).} %
Thus, QM admits a formulation which is sufficiently general
to deal with general relativistic systems.  This is a
Heisenberg formulation, in which the Heisenberg equation of
motion (\ref{hei}) is replaced by the constraint equation
(\ref{qADM}).  Clock time evolution is described in this
context by appropriate diffeomorphism invariant operators
that express the correlation between a physical variable and
the clock time \cite{time}.  These are denoted ``evolving
constant of the motion''.  An alternative explicitly
covariant generalization of QM that can deal with general
relativistic systems is Hartle's generalized QM
\cite{hartle}.  See also \cite{Ish94}.

As mentioned at the end of Section \ref{rel}, in the classical theory
the physical ``flow'' of time is presumably of thermodynamical origin. 
Remarkably, there is an intriguing quantum field theoretical analog of
(\ref{tomc}), given by a key structural property of von Neumann
algebras, the Tomita-Takesaki theorem \cite{tomita}.  As shown in
\cite{alain}, the state dependent thermal time flow can be identified
with the one-parameter group of automorphisms of the observable
algebra given by the Tomita flow of a generic state.  A state
independent characterization of the time flow is then provided by the
co-cycle Radon-Nikodym theorem \cite{connes}, which defines a {\em
canonical\/} one-parameter group of outer automorphisms of the
algebra.

The problem of merging GR and QM, and concluding a century long
scientific revolution is the problem of constructing a nontrivial {\em
weakly nonlocal\/} quantum field theory, in which locality is only
relationally defined.  To define such a theory, we need a Hilbert
space of states $\cal H$ and an operator algebra of observables $\cal
A$ on $\cal H$ that do not carry spatiotemporal dependence, but,
instead, represent diffeomorphism invariant physical states and
observables.  In the rest of this essay, I briefly illustrate several
concrete attempts to construct theories of this type.

\section{Towards general relativistic quantum field theory}
\label{sec3}

Two main avenues towards the construction of a rigorous four
dimensional general relativistic quantum field theory are
being explored: the canonical and the covariant one.  I
discuss the canonical approach in Section \ref{canonical}
and some covariant approaches in Section \ref{covariant},
below.  I then discuss the convergence of the two
approaches.  Before going into this, however, in Section
\ref{old} I briefly recall the old explorations of the
canonical and covariant approaches developed during the the
sixties and the seventies.  These explorations were very
``formal'': badly ill-defined mathematical symbols appear in
the equations, and any attempt to a nontrivial calculation
yields uncontrolled infinities.  In spite of this, the
explorations of the sixties and seventies have played an
important role in opening the way towards general
relativistic quantum field theory, because they suggested
the general structure the theory should have, and built the
physical intuition on general relativistic quantum physics. 
In a sense, many of the later developments can be seen as
efforts to transform these early constructions into rigorous
mathematics.

\subsection{Old ideas and intuitive constructions}
\label{old}

The canonical framework has been introduced by Brice DeWitt 
and John Wheeler \cite{wheelerdewitt}.  It is synthesized in 
their celebrated equations, which, in the absence of matter, 
read
\begin{equation}\label{wdw} 
\hat H\ \Psi[q] = \left[(q^{ac}q^{bd}-{\scriptstyle\frac{1}{2}} 
q^{ab}q^{cd})\ \frac{\delta}{\delta 
q_{ab}}\frac{\delta}{\delta q_{cd}}-\,\det 
q\ R[q]\right]\Psi[q]=0; 
\end{equation}
\begin{equation}\label{wdw2} 
\Psi[q] = 
\Psi[\phi(q)], \hspace{1cm} \phi\in Diff_{\Sigma}\ . 
\end{equation}
Here $q$ is the 3d metric of a spatial hypersurface 
$\Sigma$, $a, b \ldots = 1,2,3$ are tangent space indices, 
$R$ is the Ricci scalar, and the quantum state of the 
gravitational field is represented by the wave functional 
$\Psi[q]$.  Equation (\ref{wdw2}) requires $\Psi[q]$ to be 
invariant under diffeomorphisms of $\Sigma$, while the 
first, the actual Wheeler-DeWitt equation, a system of 
infinite (because $\hat H$ is a function on 3-space) coupled 
functional differential equations, is obtained as the Dirac 
quantization of the constraint that generates coordinate 
time translations in classical hamiltonian general 
relativity.  The two equations (\ref{wdw},\ref{wdw2}) can be 
seen as an implementation of 4d diff-invariance.

The interpretation of $|\Psi[q]|^{2}$ as a probability
density for a measurement of the spatial geometry to yield
the result $q$ is unfortunately common in the literature,
but is wrong, because only 4d diff invariant quantities are
observable.  To extract physical information from the
Wheeler DeWitt theory, one needs a 4d diff invariant
observable $Q[q,p]$ written in terms of $q$ and its
conjugate momentum $p$, and a corresponding quantum operator
$ \hat Q = Q[q,\frac{\delta}{\delta q}] $, commuting with
the Wheeler DeWitt operator $\hat H$.  Then the expectation
value of $Q$ in a state $\Psi$ that solves
(\ref{wdw},\ref{wdw2}) is given by $ \langle\Psi|\hat
Q|\Psi\rangle, $ where $\langle \ \ | \ \rangle$ is a scalar
product on the space of the solutions of (\ref{wdw})
determined by the requirement that the operators
corresponding to real observables be self-adjoint.

Formal solutions of (\ref{wdw}) can be obtained by using a 
covariant formalism based on Hawking's euclidean sum over 
Riemannian geometries \cite{hawk}
\begin{equation}
	Z=\int\ [Dg]\ e^{-S[g]}\ .
	\label{haw}
\end{equation}
$g$ is a Riemannian metric of a 4d compact manifold $\cal M$
and $S[g]$ is the euclidean Einstein-Hilbert action.  The
restriction of the above integration to the 4-metrics on a
4d manifold $\cal M$ bounded by a compact 3d manifold
$\Sigma$ with induced metric $q$, gives the celebrated
Hartle-Hawking's solution of (\ref{wdw}) \cite{hh}:
\begin{equation}
	\Psi[q]=\int_{\partial g=q}\ [Dg]\ e^{-S[g]}.
	\label{hawhar}
\end{equation} 
One can can consider a 4d cylinder ${\cal M}=\Sigma\times [0,1]$ and
integrate over the 4-metrics on $\cal M$ that induce the two 3-metrics
$q', q$ on the two components of its boundary.
The quantity
\begin{equation}
	P_{\Sigma}(q',q)=\int_{\partial g=q'\cup q}\ [Dg]\ e^{-S[g]}
	\label{hawpro}
\end{equation} 
is formally a projector on the solutions of (\ref{wdw},\ref{wdw2}). 
It can also be seen as a ``propagator'', from the initial to the final
$\Sigma$.  It is a propagator, however, that acts as the identity on
physical states: not surprisingly, since evolution over the
``spacetime'' manifold $\cal M$, or coordinate time evolution, is
actually pure gauge.


As already mentioned, both the canonical theory (\ref{wdw}) 
and the covariant theory (\ref{haw}) are ill 
defined and little can be computed with them.  But a number 
of well defined constructions have been inspired by these 
theories.

\subsection{Canonical approach: loop quantum gravity}
\label{canonical}
\begin{flushright}
	\begin{minipage}{11cm}
		{\em ``I feel that there will always be something 
		missing from other methods which we can only get by 
		working from a hamiltonian (or maybe from some 
		generalization of the concept of hamiltonian)''}
		
		{{\ }}\hspace{8cm} P.A.M.\, Dirac
	\end{minipage}
\end{flushright}
\vskip.2cm 

Loop quantum gravity is a well developed attempt to define a 
general relativistic quantum field theory using canonical 
methods.  It is a canonical quantization of general 
relativity, with its conventional matter couplings.  The 
theory is based on the idea of using loop variables for 
describing the gravitational field.  Loop variables have 
long been suspected to play a key role in gauge theories and 
gravity \cite{oldloop}.  The discovery of the loop 
representation is that the use of these variables turns out 
to greatly simplify the treatment of diff invariance and of 
the dynamics in quantum gravity.  In turn, the theory 
suggests that one-dimensional excitations (more precisely, 
excitations dual to surfaces in 3d) are natural diff 
invariant quantum excitations of the gravitational field.

I sketch here the basics of the formalism, focusing on its
main structure and leaving out many important details.  For
a general overview of loop quantum gravity and complete
references, see \cite{loop}.  For a pedagogical
introduction, see \cite{Karpacz}.  Let ${\cal A} = \{ A \}$
be the space of the smooth $SU(2)$ connections on a fixed
compact 3d manifold $\Sigma$.  The space $\cal A$ can be
taken as the configuration space for GR; see for instance
\cite{Abhaybook,Rovelli91,Barbero}.  The momentum conjugate
to $A$ is an $su(2)$ valued 2-form $E$, which is physically
interpreted as the densitized triad %
\footnote{I identify a vector density $E^{a}$ and the 
corresponding 2-form $ E = E^{a} \epsilon_{abc} dx^{b} 
dx^{c}$.}%
, where $(det q) q^{ab} = Tr[E^{a}E^{b}]$.  Continuous 
functionals $\Psi(A)$ form a topological vector space $L$.  
Let $U[A,\gamma]$ be the holonomy of the connection $A$ 
along the curve $\gamma$.  Let a graph 
\mbox{$\Gamma=\{\gamma_1,\ldots,\gamma_n\}$} be a finite 
collection of $n$ piecewise smooth curves, or links, 
$\gamma_i, \, i=1,\ldots,n$ in $\Sigma$, that meet, if at all, 
only at their endpoints.  Given a graph $\Gamma$ and a 
complex, Haar-integrable, function $f$ on $[SU(2)]^n$, 
consider the functional
\begin{equation}
\label{quantumstates}
\Psi_{\Gamma,f}(A) := f(U[A,\gamma_{1}], \ldots, 
U[A,\gamma_{n}])\; . 
\end{equation}
These functionals are dense in $L$.  Obviously, a functional 
based on a graph $\Gamma$ can always be rewritten as one 
based on a larger graph $\Gamma'$ that contains $\Gamma$ as 
a subgraph: it suffice to take $f$ independent from the 
holonomies of the links in $\Gamma'$ but not in $\Gamma$.  
Therefore, any two cylindrical functions can be viewed as 
being defined on the same graph.  Taking this into account, 
a scalar product is defined for any two such functionals by
\begin{equation}
\label{scalar_product}
  \left< \Psi_{\Gamma,f} |\, \Psi_{\Gamma,g} \right> := 
  \int_{[SU(2)]^n} dU_1 \ldots dU_n \: \overline{f(U_1, 
  \ldots, U_n)} \: g(U_1, \ldots, U_n) \;.
\end{equation}
$dU_1 \ldots dU_n$ is the Haar measure on $[SU(2)]^n$.  The 
scalar product (\ref{scalar_product}) is invariant under the 
natural transformation of $\Psi(A)$ under $SU(2)$ gauge 
transformations and diffeomorphisms.  The extended Hilbert 
space ${\cal H}_{ext}$ of the quantum theory is obtained by 
completion.  There exist a number of mathematical 
developments connected with the construction given above.  
They involve projective families and projective limits, 
generalized connections, representation theory of 
$C^*$-algebras, measure theoretical techniques, and others.  
See for instance \cite{Ashtekar93,Lewand_Karpacz99}.

An algebra of well defined self-adjoint field operators is defined in
${\cal H}_{ext}$.  The trace of the holonomy of $A$ around a loop 
$\alpha$, 
\begin{equation}
	T[\alpha] = {\rm Tr}\ U[A,\alpha]
	\label{eq:Top}
\end{equation}
is a multiplicative self-adjoint operator.  By replacing the conjugate
momentum $E$ (a 2-form, we recall) with a functional derivative and
integrating it over a 2d surface $C$ in $\Sigma$ we obtain a
self-adjoint Lie algebra valued operator
\begin{equation}
         E(C) = \int_{C} 
         d\sigma^{a}d\sigma^{b}
         \epsilon_{abc}\ \frac{\delta}{\delta 
         A_{c}(x(\vec\sigma))}\ \ .
\end{equation}
$T(\alpha)$ and $E(C)$ are the elementary operators of the theory, in
the same sense in which the creation and annihilation operators $a(k)$
and $a^{\dagger}(k)$ are the elementary operators in conventional
quantum field theory.  Unlikely $a(k)$ and $a^{\dagger}(k)$, the
operators $T(\alpha)$ and $E(C)$ do not require a background metric to
be defined.

The integral ${\sf A}(C)$ over the surface $C$ of the $su(2)$ norm 
of $E$
\begin{equation}
	{\sf A}(C) = \int_{C} |E|
\end{equation}
is the standard geometrical classical expression for the 
area of $C$ \cite{areaE}.  The corresponding quantum 
operator can be constructed on $H_{ext}$.  Since ${\sf A}(C)$ 
involves the square of $E$, to define it we actually need to 
regularize the classical expression and to study the limit 
of the regularized operator as the regulator is removed.  
Remarkably, the limit is finite and well defined 
\cite{RovelliSmolin95}.  The resulting operator is self 
adjoint and has discrete spectrum.  The main sequence of the 
spectrum, restoring physical units, is 
\cite{RovelliSmolin95}
\begin{equation}\label{spectrum}
	{\sf A} = 8\pi\hbar G\sum_{i=1,n}\sqrt{j_{i}(j_{i}+1)}\ , 
\end{equation}
where the $(j_{i})$ are an arbitrary n-tuplet of half
integers.  This result is the source of the loop quantum
gravity physical prediction, first suggested in \cite{area}
and in \cite{leearea}, that the area is quantized, namely
that a Planck scale sensitivity measurement of an area can
only yield discrete outputs from (\ref{spectrum}).

An orthonormal basis that diagonalizes the area operator is
given by the \emph{spin network} states
\cite{RovelliSmolin95b,Smolin97}.  Given a graph $\Gamma$,
embedded in the 3-manifold $\Sigma$, the assignment a
non-trivial irreducible representation of $SU(2)$, labeled
by its spin $j_i$, to each one of its links $\gamma_i$ is
called a {\em coloring\/} of the links.  Consider then a
node $n$ of the graph, with $k$ adjacent links $\gamma_1,
\ldots, \gamma_k$, colored as $j_{1}, \ldots, j_{k}$.  Fix
an orthonormal basis in the tensor product
$H_{n}=\otimes_{{\scriptsize i=1,k}}\: H^{(j_{i})}$ of the
Hilbert spaces of the $SU(2)$ representations $j_{1},
\ldots, j_{k}$.  The choice of an element $N_{n}$ of this
basis is called a \emph{coloring of the node $n$}.  A
(non-gauge invariant) \emph{spin network} $S=\left(\Gamma,
j_i, N_{n}\right)$ is a graph embedded in space in which
links and nodes are colored.  See Figure 1.  The \emph{spin
network state} $\Psi_S(A)$ is defined by
\begin{equation}
\label{statedef}
\Psi_S(A) = \bigotimes_{\mbox{\scriptsize links 
} i} R^{j_i}(U[A,\gamma_{i}])\ \cdot \ 
\bigotimes_{\mbox{\scriptsize nodes } 
n} N_n \; .
\label{spinnetstate}
\end{equation} 
$R^{j}(U)$ is the representation matrix of the group element
$U$ in the spin-$j$ irreducible representation of $SU(2)$,
seen here as an element in $H^{(j)}\otimes H^{(j)}$, and $\
\cdot\ $ indicates the scalar product in
$H=\otimes_{{\scriptsize{\rm links}\
i\in\Gamma}}\:(H^{(j)}\otimes H^{(j)})$.  By varying the
graph and the colors we obtain a family of states, which can
be easily normalized.  As a straightforward consequence of
the Peter--Weyl theorem, these states form an orthonormal
basis in ${\cal H}_{ext}$.

\begin{figure}
\centerline{\includegraphics[width=7cm]{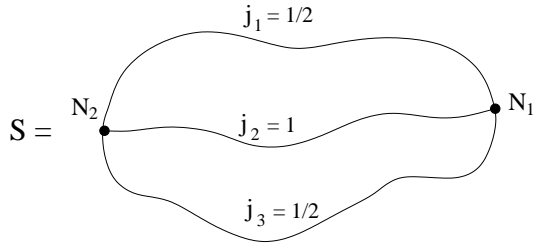}}
\centerline{\caption[]{A simple spin network with two trivalent nodes.}}
\label{spin_net1}
\label{due}
\end{figure}

If a surface $C$ cuts the links $\gamma_{1}\ldots
\gamma_{n}$ of a spin network $S$, then the spin network
state $\Psi_{S}$ is an eigenstate of the area operator with
eigenvalue (\ref{spectrum}) where the $j_{i}$'s are the spin
associated to the links $\gamma_{i}$.  See Figure 2. 
Therefore a spin network state can be thought as quantum
excitation of the metric in which each link carries a
quantum of area, proportional to the square root of the
Casimir of the representation that colors the link.  A
similar result holds for the volume \cite{RovelliSmolin95}
(see details in \cite{DePietriRovelli96}): the discrete
quanta of volume are localized on the nodes of the spin
network, and depends on the colors of the node.\footnote{The
fact that the volume operator vanishes outside the nodes can
be intuitively understood as follows: the result of the
action of the triad operator, which is a functional
derivative on the holonomy is proportional to the tangent of
the loop.  In the volume element, $\epsilon_{abc}{\rm Tr}
[E^{a}E^{b}E^{c}]$, we need at least three distinct tangents
at a point in order to have a nonvanishing triple product. 
Thus, we need a node.}.  Thus, spin network states can be
seen as elementary quantum excitations of the gravitational
field, carrying quanta of volume (``chunks of space'') at
their nodes, and quanta of area on the links that separate
the nodes, or, more precisely, on the surfaces separating
the quanta of volume, and which are dual to the links.

\begin{figure}  
\centerline{\includegraphics[width=7cm]{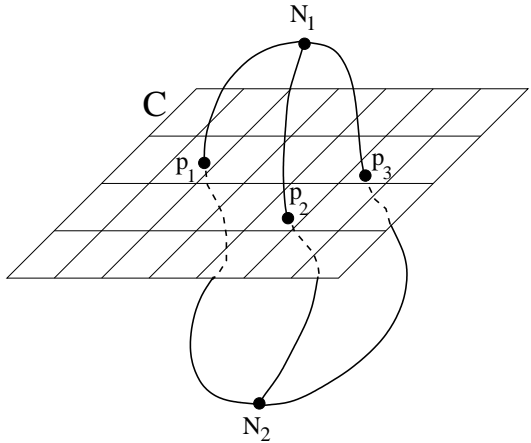}}
\centerline{
\caption[]{A spin network $S$ intersecting the surface $C$.}}
\label{spinnet_on_sigma}
\label{tre}
\end{figure}

So far, I have illustrated the kinematics of the theory. 
The next step is to implement the constraints, namely to
define the analog of Equations (\ref{wdw},\ref{wdw2}).  In
addition, we have to impose the local $SU(2)$ gauge
invariance which is peculiar to the connection formulation
of GR. $SU(2)$ local gauge transformations act on a spin
network state simply by the $SU(2)$ action on the spaces
$H_{n}$.  By restricting the colors of the nodes to basis
elements $N_{n}$ in the spin zero irreducible component of
$H_{n}$, we easily obtain the $SU(2)$ invariant subspace
${\cal H}_{0}$ of ${\cal H}_{ext}$.  Such $N_{n}$ are called
``intertwiners'', since they define invariant mappings
between $SU(2)$ representations.

${\cal H}_{0}$ carries a unitary representation $U(Diff)$ of 
$Diff_{\Sigma}$.  There are no invariant proper states, but using 
standard generalized states techniques, we can nevertheless 
define
\begin{equation}
	{\cal H}_{diff} \sim \frac{{\cal H}_{0}}{Diff_{\Sigma}}
\end{equation}
as the Diff invariant subspace of ${\cal S}'$ in a Gelfand
triple ${\cal S}' \supset {\cal H}_{0}\supset{\cal S}.$ It
is natural to take ${\cal S}$ as the space of the finite
linear combinations of spin network states.  The spin
network basis in ${\cal H}_{0}$ yields a simple description
of ${\cal H}_{diff}$ and a definition of the Hilbert
structure on it.  Let an $s$-knot, or abstract spin network,
$s$ be an equivalence class under $Diff(\Sigma)$ of
(embedded) spin networks $S$.\footnote{In fact, by slightly
enlarging $Diff_{\Sigma}$ we can eliminates a moduli space
structure in the equivalence classes of the nodes with
intersections \cite{moduli}, and obtain a {\em separable\/}
Hilbert space $H_{diff}$.  On this, see also \cite{zap}. 
(The unconstrained space state $H_{ext}$ is non-separable.)}
A basis in ${\cal H}_{diff}$ is given by the states $\langle
s |$ in ${\cal S}'$ defined by
\begin{equation} 
\label{s_scalar1} 
  \langle s | S \rangle = \left\{
		\begin{array}{c@{\qquad}l}
				0 & \mbox{if\quad} S \not\in s \\
				 1 & \mbox{if\quad} S \in s \;.  \end{array} 
				 \right.
\end{equation} 
where we have used a Dirac bra-ket notation for the elements
of ${\cal S}'_{0}$ and ${\cal S}$.  Furthermore, one can
prove (see \cite{Lewand_Karpacz99} and references therein)
that observables on ${\cal H}_{0}$ which are self-adjoint
(and thus correspond to real classical quantities) and diff
invariant (and thus are well defined on ${\cal H}_{diff}$)
are self-adjoint under the scalar product
\begin{equation}
\label{s_scalar2}
  \langle s | s' \rangle = \left\{
		\begin{array}{c@{\qquad}l}
				0 & \mbox{if\quad} s \neq s' \\
				 c(s) & \mbox{if\quad} s = s \;.  
				 \end{array} \right.
\end{equation} 
where $c(s)$ is the number of discrete symmetries of the abstract
$s$-knot.  This is therefore the appropriate scalar product we need on
physical grounds, picked up by the requirement that real classical
quantities become self-adjoint operators. 

The states $|s \rangle$ are (3d) diffeomorphism invariant quantum
states of the gravitational field.  They are labeled by $s$-knots:
abstract, non-embedded, knotted, colored, graphs $s$.  As we have seen
above, each link of the graph can be seen as carrying a quantum of
area, and each node quanta of volume.  An $s$-knot represents an
elementary quantum excitation of space: its nodes represent ``chunks''
of space with quantized volume, separated by ``elementary surfaces'',
dual to the links, with quantized area.  The key point is that an
$s$-knots does not live on a manifold.  It is not ``located
somewhere''.  It is not a quantum excitations {\em in\/} spacetime. 
Rather, it is a quantum excitations {\em of\/} spacetime.  The
quantized space does not reside ``somewhere'': it itself defines the
``where''.  This is the picture of quantum space that emerges from
loop quantum gravity.  It is profoundly different from the structure
of the states of a weakly local quantum field theory.

The last step in the definition of the theory is the 
construction of the quantum hamiltonian constraint, namely 
the rigorous analog of the Wheeler DeWitt equation 
(\ref{wdw}).  This is obtained by promoting the classical GR 
hamiltonian constraint, which in the connection formalism 
can be written \cite{Thiemann96} as 
\begin{equation}
 H[N]= \int N \ {\rm Tr}[F\wedge \{A, V\}]
\label{hamiltonian}
\end{equation}
into a quantum operator.  Here $N$ is a scalar smearing
function; $F$ is the curvature of $A$; $\{\ , \}$ are
Poisson brackets; and $V$ is the volume of $\Sigma$.  To
promote $H[N]$ to an operator, we need first to regularize
it.  We replace the classical expression (\ref{hamiltonian})
by a regularized, $\epsilon$ dependent one,
$H_{\epsilon}[N]$, written in terms of quantities that we
know how to promote to quantum operators.  In particular,
$F$ is replaced by the holonomy of an $\epsilon$-size loop. 
The classical quantities are then replaced by the
corresponding quantum operators, leading to the Hamiltonian
operator $\hat{H}_{\epsilon}[N]$.  Finally, we study the
limit $\epsilon \rightarrow 0$: $\hat{H}_{\epsilon}[N]
\rightarrow \hat{H}[N]$.  This can be done in detail
\cite{Thiemann96}, with surprising results.

First, the action of the operator vanishes on the holonomy
$U[A, \gamma]$ anytime the smearing function $N$ is zero on
the end points of $\gamma$ \cite{RovelliSmolin90}.  In other
words, $\hat H$ acts only at the nodes of the spin networks. 
This can be seen as a consequence of the presence in
(\ref{hamiltonian}) of the volume.  Indeed, as described
above, the volume operator vanishes outside the nodes.

\begin{figure} 
\centerline{\includegraphics[width=7cm]{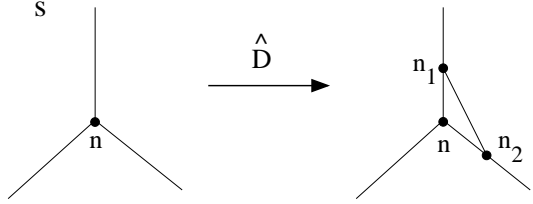}}
\centerline{ 
\caption[]{The action of the Hamiltonian constraint on a 
trivalent node.}}
\label{hamonvertex}
\label{quattro}
\end{figure}

Second, the result of the action of the hamiltonian operator 
on a spin network state turns out to be given by
\begin{equation}
\label{H_on_s}
  \hat{H}[N] \, | S \rangle = \sum_{\mbox{\scriptsize 
  nodes}\, n\, \mbox{\scriptsize of}\, s} N(x_n)\,  A_n \,
  \hat D_{n}\,   | S \rangle \;.
\end{equation}
$x_n$ is the point in which the node $n$ is located.  The
action of the operator $D_{n}$ on the state $|S \rangle $ is
given in Fig.~\ref{hamonvertex}: the operator acts by
creating an extra link which joins two points $n_1$ and
$n_2$ lying on distinct links adjacent to the node $n$
\cite{outline}.  (A sum over all couples of adjacent links
is understood.  The extra triangular loop that the new link
forms is essentially the $\epsilon$-size loop whose holonomy
gives the regularization of the curvature $F$ in
(\ref{hamiltonian}).  The coefficients $A_{n}$ can be
explicitly computed \cite{Borissov97}.

Now, the key point of this construction is that $\hat{H}[N]$ 
is well defined on ${\cal H}_{diff}$.  More precisely, the 
limit in which the regulator is removed makes sense and is 
{\em finite\/} on the diffeomorphisms invariant states, and 
only on those states.  It is here that one sees the deep 
interplay between general covariance and quantum field 
theory.  What happens is that the size of the triangle in 
Figure 3, or, equivalently, the precise 
position of the two new nodes, depends on the regulator 
$\epsilon$.  However, since the action of the operator on a 
state $\langle s |$ in ${\cal H}_{diff}$ is defined by 
duality, the relevant limit is
\begin{equation}
\langle \hat H[N] s |S \rangle \equiv \lim_{\epsilon\to 
0}\ \langle s |\hat H_{\epsilon}[N] S \rangle. 
\label{limit}
\end{equation}
But when acted upon by the diff invariant state $\langle s
|$, the precise position of the link in $|\hat
H_{\epsilon}[N] S \rangle$ is irrelevant, because different
positions are related by a diffeomorphism, for $\epsilon$
sufficiently small.  Therefore the limit (\ref{limit}) turns
out to be trivial (constant) \cite{RovelliSmolin94}.  In a
precise sense, potential ultraviolet divergences are washed
away by diffeomorphism invariance.

This concludes the description of the theory.  The theory
has been applied, for instance, in black hole physics,
leading to a finite computation of the Bekenstein-Hawking
entropy of a Schwarzschild black hole \cite{blackhole}. 
Several aspects of the theory are still poorly understood
(low energy limit, general observables).  Furthermore, the
general lines of the construction of the hamiltonian
constraint are understood, but a number of variants are
possible, reflecting rather wide quantization ambiguities,
and it is not clear which variant, if any at all, yields the
physically correct theory, with the right low energy limit. 
Still, loop quantum gravity provides a construction of a
well defined and nontrivial general relativistic quantum
field theory.

\subsection{Covariant approaches}
\label{covariant}

\subsubsection{Topological quantum field theories and the 
Turaev Viro model} \label{TuVi}

Well defined, although very simple, examples of a covariant
formulation of general relativistic quantum field theory are
provided by topological quantum field theories (TQFT). 
Topological field theories are theories in which the number
of gauges matches the number of fields, so that all local
degrees of freedom are washed away by gauge invariance.  If
the space topology is non-trivial, a finite number of global
degrees of freedom may remain gauge invariant, leaving a
non-trivial dynamical theory.

A prejudice hard to die wants any diffeomorphism invariant
field theory to be topological, in the sense of having a
finite number of global degrees of freedom, unless
diffeomorphism invariance is broken.  This is wrong, and, as
far as we know from GR, the world is described precisely by
a diffeomorphism invariant field theory in which
diffeomorphism invariance is not broken\footnote{Of course,
we may gauge fix it.}, but the number of degrees of freedom
is infinite.  These degrees of freedom are ``local'' in the
relational sense: they are localized with respect to each
other.  Indeed, we can distinguish three kinds of theories. 
First, there are non-diff-invariant theories like QED. Thy
have weakly local degrees of freedom --weaves-- interpreted
as localized in spacetime (with respect to a reference
system).  Second, there are theories like GR. These are
diffeomorphism invariant, have an infinite number of degrees
of freedom --gravitational waves--.  These degrees of
freedom are relationally local and weakly nonlocal (hence
the long historical confusion, now completely over, on
whether gravitational waves are ``real'' or not).  Finally,
there are topological theories, which are diffeomorphism
invariant, and, in addition, have only a finite number of
global degrees of freedom.

GR, on the other hand, {\em is\/} topological in three
spacetime dimensions (there are no gravitational waves in
3d).  Quantum GR in 3d has been intensively studied
\cite{witten,carlip,3dnoi}.  A state sum formulation of a
TQFT that yields euclidean 3d GR in an appropriate limit is
provided by the celebrated Turaev-Viro model
\cite{Turaev-Viro}.  The Turaev-Viro model can be defined
over an arbitrary triangulation $\Delta$ of the spacetime
3-manifold.  An assignment of an irreducible representation
$j_{e}$ of the quantum group $SU(2)_{q}$ to each 1-simplex
(edges) $e$ of the triangulation is called a coloring $c$ of
the triangulation.  We chose $q$ a root of unity so that
that the number of irreducible representations is finite. 
The partition function of the model is defined as a sum over
all colorings of $\Delta$ by
\begin{equation}
    Z_{TV} = \sum_{c} \ \prod_{e} {\rm dim}(j_{e}) 
	\ \prod_{t} \{6j\}_{t}(c). 
    \label{tv}
\end{equation} 
${\rm dim}(j_{e})$ is the (quantum) dimension of the
representation $j_{e}$.  The second product is over the
tetrahedra $t$ of the triangulation.  $\{6j\}_{t}(c)$ is the
$q$-analog of the Wigner 6-j symbol of the six
representations assigned to the edges that bound the
tetrahedron $t$.  The Wigner 6-j symbol of $SU(2)$ can be
computed as the value, for $A=0$ of the spin network state
(\ref{spinnetstate}) where the spin network is given by the
colored one skeleton of $t$.  The analogue calculation for a
quantum group can be done using the Temperley-Lieb
recoupling theory techniques developed in \cite{kl}.  The
sum (\ref{tv}) turns out to be independent from the
triangulation.  All triangulations are connected by a small
set of elementary moves, and (\ref{tv}) does not change if
we perform an elementary move on the triangulation.  The
invariance of (\ref{tv}) under these moves follows from the
properties of the the Wigner 6-j symbols.  In turn, these
reflect the fact that the irreducible representations form
an {\em associative\/} tensor category \cite{crane}.

The Turaev-Viro model can be defined over a triangulated 
3-manifold {\em with boundary}.  We associate a Hilbert 
space $H_{\Sigma}$ to each component $\Sigma$ of 
the boundary of the 3-manifold $\cal M$.  This is done as 
follows.  First, associate to $\Sigma$ a Hilbert space 
$L_{\Sigma}$, spanned by an orthonormal basis of states 
$|s\rangle_{\Sigma}$, where $s$ is a coloring of the 
triangulation of $\Sigma$.  Then, consider a 3-manifold 
bounded by $\Sigma\cup\Sigma'$.  Define a linear map between 
$L_{\Sigma}$ and $L_{\Sigma'}$ with matrix elements
\begin{equation}
	P_{\Sigma,\Sigma'}(s,s') = {}_{\Sigma}\langle 
	s|P|s'\rangle_{\Sigma'} = \sum_{c, \partial c = s\cup 
	s'} \ \prod_{t} \{6j\}_{t}(c).
    \label{tvb}
\end{equation}
The sum is over the coloring of the internal edges 
only.  If $\Sigma$=$\Sigma'$ and $\cal M$ is the cylinder 
$\Sigma\times[0,1]$, (\ref{tvb}) defines a projector 
\begin{equation}
	P_{\Sigma}(s,s') =	P_{\Sigma,\Sigma}(s,s') 
	    \label{tvb2}	
\end{equation}
on $L_{\Sigma}$.  $H_{\Sigma}$ is the kernel of this 
projector, and is triangulation independent.

In general, the map (\ref{tvb}), restricted as a map from
$H_{\Sigma}$ to $H_{\Sigma'}$ is independent from the
triangulation.  In fact, this construction defines a functor
from the category $\sf M$ whose objects are 2-manifolds and
morphism are 3-manifolds with boundaries, into the category
$\sf H$ of Hilbert spaces.  From this point of view, the
projection
\begin{equation}
	P : L_{\Sigma} \longrightarrow H_{\Sigma}
	\label{projector}
\end{equation}
defined by (\ref{tvb2}) is the standard technique to get a
functor from a semifunctor.  Atiyah has provided a
compelling general definition of TQFT as a functor from $\sf
M$ to $\sf H$ \cite{Atiyah}.  Atiyah's general scheme
captures the structure we expect from a general relativistic
quantum field theory, and, in particular, from a quantum
theory of gravity.  Notice that the projection
(\ref{projector}) provides precisely the analog of the the
Wheeler DeWitt equation.  In particular, it represents a
realization of Hawking's projection (\ref{hawpro}) on the
solutions of (\ref{wdw}), which is also defined on a
cylinder.  The point is that evolution along the cylinder
$\Sigma\times[0,1]$ is the evolution in coordinate time
generated by the hamiltonian constraint.  This evolution, I
recall, is a non-physical gauge transformation in the
classical theory.  The analogy between quantum gravity and
TQFT goes much further than this formal similarity
structure: as we shall see in Section \ref{spinfoams}, the
solution of the loop quantum gravity version of the
Wheeler-DeWitt equation, can be computed by a formula
surprising similar to (\ref{tvb}).

On the other hand, the axioms of a TQFT in Atiyah's
formulation require the spaces $H_{\Sigma}$ to be finite
dimensional.  It is this feature that reflects the
topological nature of the theories, and is not compatible
with quantum gravity, where we expect the theory to have an
infinite number of degrees of freedom and an infinite
dimensional state space.  The rest of the formal structure
of Atiyah's definition of TQFT, on the other hand, reflects
only the diffeomorphism invariance of the theory, and it is
thus likely to underlie a full quantum theory of gravity as
well.

Now, remarkably, in the Turaev-Viro model the states in
$L_{\Sigma}$ have a natural representation as spin network
states.  This is the first of a number of structural
similarities between TQFT and loop quantum gravity.  To see
this, consider the {\em dual\/} $\Delta'$ of the
triangulation $\Delta$.  Going to the dual triangulation
will be a crucial technique below \cite{fotini}.  $\Delta'$
is a cellular complex whose 1-cells correspond to the
triangles in $\Delta$ and whose 2-cells correspond to the
edges in $\Delta$.  The coloring of $\Delta$ induces a
coloring of the 2-cells of $\Delta'$.  Consider now a
connected component $\Sigma$ of the boundary.  The dual of
the triangulation of $\Sigma$, or, equivalently, the
boundary of $\Delta'$, is a trivalent graph.  A state,
namely a coloring of $\Sigma$ determines a coloring of the
links of this graph.  This is precisely a (trivalent) spin
network.  A link carrying the trivial representation is
viewed as a non existing link.  Recall that $L_{\Sigma}$ is
spanned by the coloring of the triangulation of the
boundary: it follows that $L_{\Sigma}$ admits a basis of
spin network states.

In the $q=1$ case the model diverge because there is an
infinite number of representations.  The divergent sum is
the Ponzano-Regge model, a formal quantization of 3d GR
constructed in the late sixties \cite{Ponzano:1968}), which
has inspired most of the later developments.  The $q=1$
model can be also seen as a quantization of a 3d topological
field theory, $BF$ theory \cite{BF}.  The fields of $BF$
theory are a $SU(2)$ connection $A$, and a $su(2)$-algebra
valued 1-form $B$.  The action is
\begin{equation}
	S[A,B]=\int {\rm Tr}[B\wedge F].
	\label{bf}
\end{equation}
A discretization and path integral quantization of this 
theory yields (\ref{tv}).  Because of the topological aspect 
of the theory, the discretization turns out not to change 
the theory itself; that is, no degrees of freedom are lost 
in the discretization, or, equivalently, the continuum limit 
of the discretization is trivial.  A canonical quantization 
\cite{witten} yields the same quantum theory, and, 
remarkably, one finds that the spin network states are 
precisely represented by spin network functionals of the 
connection \cite{oogurirovelli,foxon}.

\subsubsection{Four dimensions: the Turaev-Ooguri-Crane-Yetter model}

Let us now move to four dimensions.  The generalization of
the Turaev-Viro model to four dimensions was found by
Turaev, Ooguri and Crane and Yetter
\cite{Turaev,Ooguri:1992b,Crane-Yetter,Crane}.  The key idea
is to color 2-simplices (faces) $f$ of the triangulation
with irreducible representations $j_{f}$ and to color
3-simplices (tetrahedra) $t$ with intertwiners $N_{t}$.  One
then defines
\begin{equation} 
    Z_{CY} = \sum_{c}  \ \prod_{f} {\rm dim}(j_{f}) 
	\ \prod_{s}\ \{15j\}_{s}(c)
    \label{cy}
\end{equation}
where the second product is now over the 4-simplices $s$ of
the triangulation, and the Wigner 6-j symbol is replaced by
the 15-j symbol.  The Wigner 15-j symbol can be computed as
follows.  In the {\em dual\/} triangulation, the 2 and 3
simplices that bound $s$ correspond to 2-cells and 1-cells. 
The intersection of these with a ball surrounding $s$ (which
in the dual triangulation is a point) is again a spin
network.  The 15-j symbol is the $A=0$ value of the
corresponding spin network state.  The
Turaev-Ooguri-Crane-Yetter (TOCY) model is the quantization
of 4d $BF$ theory.  In 4d $BF$ theory the actions is as in
(\ref{bf}), but $B$ is now a two form.

In Section \ref{TuVi}, I have shown that the states of the 
Turaev-Viro model can be described as spin networks: a basis 
of states in the (unconstrained, that is, before the 
projection) state space $L_{\Sigma}$ is given by the colored 
1-skeletons of the cellular complex dual to the boundary 
$\Sigma$, and this is precisely a spin network.  The same is 
true for the TOCY model.  Indeed, the 
boundary $\Sigma$ is now three-dimensional; the colored 
triangulation of $\Sigma$ carries spins over its triangles 
and intertwiners over its tetrahedra. Its dual carries spins 
over its edges and intertwiners over its nodes.  These data 
define precisely a spin network. Therefore spin networks 
label states in 3 as well as in 4 dimensions. 

Now, the euclidean GR action can be written as 
\begin{equation}
	S[A,E]=\int {\rm Tr}[E\wedge E\wedge F].
	\label{graction}
\end{equation}
where $A$ is an $SO(4)$ connection and $E$ the (inverse)
tetrad field, which can be seen as a $SO(4)$ algebra valued
1-form.  Comparing with (\ref{bf}), we see that GR can be
considered as an $SO(4)$ $BF$ theory plus a suitable
constraint, imposing, essentially, the two form $B$ to be
the exterior product of two 1-form fields $E$ \cite{DF}.  It
was then natural to search for a formulation of quantum
euclidean GR as a (non topological) modification of a
$SO(4)$ TOCY state sum model.  There are several attempts to
do that, leading to some tentative formulations of quantum
euclidean GR as state sums of the form (\ref{cy}).  In
particular, Reisenberger \cite{Reisenberger} has proposed to
directly implement the constraint on appropriate tensor
products of the Hilbert spaces on which the irreducible
representations are defined.  The constraint has an
appealing geometrical interpretation as conditions for
triangles defined by $B$ (in 4d) to join into tetrahedra
\cite{Barbieri}.  Barrett and Crane \cite{Barrett:1998}
noticed that these conditions, in turn, admits a direct
interpretation in terms of $SO(4)$ representation theory:
they are implemented within an ${\rm SO}(4)$ Crane-Yetter
state sum model simply by (appropriately) restricting the
sum to {\em simple\/} $SO(4)$ representations.\footnote{%
Irreducible representations of $SO(4)$ are labeled by two
half integers (two spins) $(j',j")$ where $j'+j''$ is
integer.  A representation is simple if $j'=j"$.  Thus,
simple representations are labeled by just one spin
$j=j'=j"$.} A number of results support the idea that the
these models model are indeed related to quantum GR
\cite{altridue,FW}.

The key problem in these models is the following.  In the
topological theories, the introduction of a fixed
triangulation is justified by the triangulation independence
of the quantum theory.  Triangulation independence reflects
the topological aspect of the theory and the absence of the
local degrees of freedom: simply speaking, a finite number
of variables suffice to capture all the physical degrees of
freedom of the theory.  In modifying $BF$ theory to obtain
GR, one looses triangulation independence.  This is to be
expected, since GR has infinite relationally local (although
weakly nonlocal) physical degrees of freedom.  In this
context, a triangulation represents a genuine cut off of the
degrees of freedom.  Thus, the theory defined over a fixed
triangulation cannot be the physical theory one is
searching.  To get the physical theory, we have either to
``sum over all triangulations'', or to take a limit ``in
which the triangulation becomes finer and finer''.  These
limits are poorly understood.  Some control over a sum over
the sum over triangulations, however, can be obtained from
an alternative approach to these models, which I illustrate
in the next section.

\subsubsection{Nonperturbative string theory in 0 
dimensions, field theories over a group and summing over 
triangulations}

An intriguing non perturbative and genuinely background 
independent formulation of 2d quantum gravity was 
obtained sometime ago in the context of string theory ``in 
zero dimensions''.  This is the theory obtained by dropping 
the scalar fields on the string world sheet that represent 
the location of the string in the target space, and 
retaining just the 2d metric as a dynamical variable.  The 
resulting theory can be expressed a sum over the geometries 
of a two dimensional surface, as in (\ref{haw}).  This sum 
can be concretely defined by triangulating the 2d surface 
with triangles with sides of fixed length, and summing over 
all topologically distinct triangulations.  The number of 
triangles joining on a vertex determines the local curvature 
of the manifold at the vertex.  Remarkably, the sum over 
such triangulations can be reinterpreted as the Feynman 
expansion of the partition function of a quantum theory of 
matrices with a simple cubic potential \cite{2d}.  Indeed, 
the {\em dual\/} of a triangulation is a trivalent graph, 
and can be seen as one of the trivalent Feynman graphs of 
the cubic matrix theory.

A remarkable extension to 3d of the idea of these matrix 
models was obtained by Boulatov in \cite{Boulatov:1992}.  
Boulatov considers a field theory over three copies of 
${\rm SU}(2)$, with the action
\begin{equation}
	S[\phi]=
	\int dg_{1}dg_{2}dg_{3} \ (\phi(g_{1},g_{2},g_{3}))^{2}+
	\frac{\lambda}{4!} \int dg_{1}\ldots dg_{6} \ 
	\phi(g_{1},g_{2},g_{3})
	\phi(g_{3},g_{4},g_{5})
	\phi(g_{2},g_{4},g_{6})
	\phi(g_{1},g_{5},g_{6}). 
	\label{boulatov}
\end{equation}
Notice that if we represent the field by a vertex and its 
three arguments with three edges attached to this vertex, 
then the second term in the action has the structure of a 
tetrahedron.  Now, if we expand $\phi$ in modes, harmonic 
analysis on $SU(2)$ teaches us that the ``momenta'' of 
the field $\phi$, that is, its modes, are labeled by the 
irreducible representations of the group.  If we express 
(\ref{boulatov}) in terms of these modes and then compute 
the Feynman's perturbative expansion in $\lambda$ of the 
partition function 
\begin{equation}
   Z = \int [D\phi]\ e^{iS[\phi]} = \sum_{\Gamma} \ 
   \lambda^{n(\Gamma)}\  Z_{B}[\Gamma], 
\end{equation}
we obtain Feynman graphs $\Gamma$ (with $n(\Gamma)$ 
vertices) formed by tetravalent vertices connected by 
propagators.  Each propagator has three indices, carrying 
momenta, namely $SU(2)$ irreducible representations.  These 
indices are paired across the vertex, and summed over in 
such a way that they define closed loops, or cycles, along 
the graph.  If we interpret these cycles as defining 
2-cells, the Feynman graph is a 2-complex with 2-cells 
colored by $SU(2)$ irreducibles.  Most remarkably, if the 
2-complex $\Gamma$ is the dual 2-skeleton $\Gamma(\Delta)$ 
of a triangulation $\Delta$, then the sum over momenta
\begin{equation} 
    Z_{B}[\Gamma] =  \sum_{c} A[\Gamma,c] 
\end{equation}
is precisely the Turaev-Viro sum over colorings of $\Delta$, with
$q=1$.  That is, the possible momenta on $\Gamma(\Delta)$ match exactly
the possible colorings on $\Delta$, and each Feynman amplitudes is
equal to the corresponding Turaev Viro amplitude.
\begin{equation} 
A[\Gamma(\Delta),c] = \prod_{e}\ {\rm dim}(j_{e}) \prod_{t}\
\{6j\}_{t}(c).
\end{equation}
Therefore, as formal series 
\begin{equation} 
     Z_{TV}[\Delta] = Z_{B}[\Gamma(\Delta)] \ . 
\end{equation}

The same trick works in four dimensions, as realized by 
Ooguri \cite{Ooguri:1992b}.  The Ooguri theory is a field 
theory over 4 copies of ${\rm SU}(2)$. Its action has the 
structure
\begin{equation}
	S[\phi]=
	\int dg_{1}\ldots dg_{4}\  \ \phi^{2}\ +\ 
	\frac{\lambda}{5!} \int dg_{1}\ldots dg_{10}\  \ 
	\phi^{5}, 
	\label{ooguri}
\end{equation}
where the potential term has now the structure of a 
4-simplex.  The Feynman expansion of this theory determines 
a state sum for a triangulated 4d spacetime, which is 
precisely the $q = 1$ case of the TOCY state 
sum (\ref{cy}).  Again, the theory is, formally, 
triangulation independent.

Now, the modification introduced into the $SO(4)$ 
TOCY model in order to obtain GR from BF 
theory can be implemented directly in the above model 
\cite{DKFR}.  In 
fact, it is sufficient to replace $SU(2)$ with $SO(4)$ in 
(\ref{ooguri}) {\em and\/} to constrain $\phi$ to be 
invariant under a fixed $SO(3)$ subgroup $H$ of $SO(4)$:
\begin{equation}
	\phi(g_{1},g_{2},g_{3},g_{4}) = 
	\phi(h_{1}g_{1},h_{2}g_{2},h_{3}g_{3},h_{4}g_{4}),
	\hspace{1cm} \forall h_{i}\in H. 
\end{equation}
Remarkably, this constraint implements precisely the
restriction of the modes to the simple representations of
$SO(4)$, which, in turn, is the quantum implementation of
the constraint that reduces GR to BF. Given a triangulation
$\Delta$ with a dual 2-complex $\Gamma(\Delta)$, the value
of the Feynman graph $\Gamma$ is {\em precisely\/} the
Barret-Crane partition function over the triangulation
$\Delta$.  But unlikely in the previous cases, which were
topological, and therefore triangulation independent, this
time we gain something crucial: the full Feynman expansion
of the theory defines precisely a ``sum over
triangulations'', restoring full general relativistic
invariance.

The terms in the expansion can be interpreted as a
``quantized spacetime geometry''.  In fact, the
interpretation of the links and nodes of the spin network
states as carriers of quanta of area and volume extends, as
we shall see in the next section, to the spacetime colored
triangulations.  Thus these models provides a concrete
definition of Hawking's sum over geometries.  The quantum
geometries can be generated as Feynman graphs, as in the 2d
quantum gravity models.  Thus, in this context, we can view
a triangulated spacetime as a term in a Feynman expansion. 
Alternatively, we can interpret the field $\phi$ (more
precisely, its ``Fourier'' transform) as a quantum amplitude
for the geometry of a tetrahedron (see \cite{Barbieri}). 
Then the quantum field theory can be seen as a
``multiparticle theory'' for many ``tetrahedra'', or
elementary chunks of space, being created and destroyed in
interactions.  A spacetime is a Feynman history of creations
and destructions of chunks of space.

\subsubsection{Spin foam models}\label{spinfoams}

All the models described above admit a common description as
{\em spin foam models}.  This is a surprising and remarkable
fact.  Let me sketch the general structure of a spin foam
model.  A spin foam model is a Feynman-like sum over
histories, in which the histories are spin foams.  A spin
foam $\sigma$ is a colored 2d complex.  A 2d complex is an
(abstract) collection of ``faces'', which join at ``edges'',
which, in turn, join at ``vertices''.  Two 2-complexes are
considered distinct if they are combinatorially distinct.  A
coloring $c$ of a 2-complex is an assignment of spins to the
faces and intertwiners to the edges.  A spin foam model is
determined by choosing a vertex amplitude $A_{\nu}(c)$, a
function of the colorings of faces and edges adjacent to the
vertex.  The spin foam model is then defined as the state
sum
\begin{equation}  
\label{sumoverfoams} 
 Z \ \ =\  \sum_{\mbox{\scriptsize spin} 
 \atop \mbox{\scriptsize foams} \, 
 \scriptstyle \sigma} \ \ \prod_{\mbox{\scriptsize vertices} 
 \atop \scriptstyle v \in \sigma}\  A_{v} \ , 
\end{equation}
or, possibly, including amplitudes of edges and faces as well: 
\begin{equation}  
 Z \ \ =\  \sum_{\mbox{\scriptsize spin} 
 \atop \mbox{\scriptsize foams} \, 
 \scriptstyle \sigma} \ \ \prod_{\mbox{\scriptsize faces} 
 \atop \scriptstyle f \in \sigma}\  A_{f}   \ 
 \prod_{\mbox{\scriptsize edges} 
 \atop \scriptstyle e \in \sigma}\  A_{e}  \ 
 \prod_{\mbox{\scriptsize vertices} 
 \atop \scriptstyle v \in \sigma}\  A_{v} \ .
\end{equation}
If $\Delta$ is a triangulation of a manifold
$\cal M$, and $\Delta'$ the corresponding dual complex, then
the 2-skeleton of $\Delta'$ is a 2d complex, which we
indicate as $\sigma(\Delta)$.  The vertices of
$\sigma(\Delta)$ correspond to the the n-simplices of
$\Delta$; the edges to the n-1 simplices and the faces to
the n-2 simplices.  Notice that in the 3d models described
above, spins were carried by the edges of $\Delta$, while in
the 4d models spins were carried by the triangles of
$\Delta$: in both cases, we obtain a spin foam when using
the dual description.  (In the Turaev-Viro model, edges are
trivalent and thus intertwiners are unique, because there is
a single normalized invariant tensor between three $SU(2)$
representations).  Furthermore, in all models described
above, the state sum does not depend on the full
combinatorial data of the triangulation, but only on the
structure of its n, n-1, and n-2 simplices: that is to say,
just on the data that are represented by the corresponding
spin foam.  Thus, for instance, we obtain the Turaev-Viro
model by choosing $A_{v}$ to be the Wigner 6-j symbol on
the vertices that have the structure of the dual of a
tetrahedron, and to vanish otherwise.  We obtain the TOCY
model by choosing $A_{v}$ to be the Wigner 15-j symbol on
the vertices that have the structure of the dual of a
4-simplex, and to vanish otherwise; and so on.

Surprisingly, however, spin foams come from the canonical
quarters.  Spin foams were first introduced to describe the
dynamics of loop quantum gravity, under the name branched
colored surfaces, \cite{Reisenberger}.  As I illustrate
below, indeed, the covariant formulation of loop quantum
gravity yields a partition function which is again a spin
foam model.  The idea that quantum spacetime could be
described in terms of a sum over surfaces had been proposed
earlier, in particular by Baez \cite{Baez:1994}, and then by
Reisenberger \cite{Reisenberger}, and Iwasaki
\cite{Iwasaki}, who realized the importance of 2-complexes. 
The general notion and the term ``spin foam model'' were
introduced by Baez \cite{Baez}.  The possibility of defining
a causal structure over spin foams has also been explored
\cite{othersn}.  See \cite{Baez} and \cite{Baez2} for
details and full references.

Now, recall that in all the models described above,
irrespectively of the dimension, a basis of (unconstrained)
states is labeled by spin networks.  Consider the state sum
$P_{\Sigma}(s,s')$ defined in (\ref{tvb},\ref{tvb2}) over
the manifold $\Sigma\times [0,1]$.  As discussed, this
quantity defines an evolution in coordinate time, as well as
the projector on the physical state space.  Now, a moment of
reflection will convince the reader that in the spin foam
formulation, $P_{\Sigma}(s,s')$ has a remarkable
interpretation: it is a sum over ``histories'' $\sigma$ of
evolutions of the spin network state $s'$ into the spin
network state $s$.  That is to say, a spin foam can be seen
as the spacetime world-sheet of a spin network.  The faces
of the spin foam are the world-sheets swept by the links of
the spin network and the edges of the spin foam are the
worldlines swept by the nodes of the spin network.  Vertices
are where dynamics happen, precisely as in Feynman diagrams. 
That is to say, a spin foam vertex can be seen as a
``vertex'' in the sense of the Feynman graphs of
conventional quantum field theory.  A spin foam with $n$
vertices represents thus a transition amplitude across n-1
intermediate states, or an n-order term is a perturbative
expansion of the transition amplitude.  

In the Turaev-Viro theory, for instance, the vertex is the
dual of a tetrahedron.  Thus, it has four adjacent edges. 
Consider one of these edges as coming from the past, and
three going into the future.  Then the vertex represents a
process in which a trivalent node of the spin network opens
up in three trivalent nodes.  See Figure 4.  A hamiltonian
that could generate such a vertex must thus have precisely
the action described in Figure 3, and the vertex amplitude
can be seen as a matrix element of this hamiltonian between
the initial and final state.  But Figure 3 represents the
action of the hamiltonian constraint of canonical quantum
gravity, which was derived by starting from the ADM
constraint and promoting it to an operator!  This is a
remarkable convergence.
\begin{figure}
\centerline{\includegraphics[width=3.5cm]{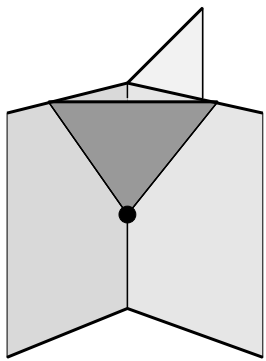}}
\centerline{\caption[]{The elementary vertex.}}
\label{vertex}
\label{cinque}
\end{figure}

I now sketch the formal derivation of the spin foam
formulation of loop quantum gravity from the canonical
theory \cite{ReisenbergerRovelli97}.  Once more, I give here
only a very sketchy account of the derivation, and refer to
the literature for all details.  The problem in canonical
quantum gravity is to define the physical Hilbert space
${\cal H}_{phys}$, which is the space of the solutions of
the hamiltonian constraint, starting from the diffeomorphism
invariant Hilbert space ${\cal H}_{diff}$ spanned by the
$s$-knot states.  Consider the ``projector'' operator
\begin{equation}
\label{P}
   P = \int [dN] \ \  e^{i \hat{H}[N]} \  \;.
\end{equation}
to be compared with the expression of Dirac's delta as the
integral of an exponential.  A diffeomorphism invariant
notion of integration exists for this functional integral
\cite{ReisenbergerRovelli97,Thiemanint}.  Consider the
matrix elements of $P$ in the $s$-knots basis, and expand
the exponent.  The expansion has the structure
\begin{equation} 
  \langle s | P | s' \rangle \sim \langle s | s' \rangle + 
  \int [dN] \, \left( \langle s | \hat{H}[N] | s' \rangle\ + 
  \langle s | \hat{H}[N] \hat{H}[N] | s' \rangle\ + \ldots \right) 
  \;.
  \label{expansion}
\end{equation}
Using now the action (\ref{H_on_s}) of $\hat{H}$ on spin
network states, we can compute explicitly the action of the
$\hat H[N]$ operators.  The resulting integrals of the type
$ \int [dN] \, \left( N(x_1) \cdots N(x_n) \right) $ can be
integrated explicitly, leaving a sum of terms, each
corresponding to a sequence of actions of the hamiltonian
constraint over an $s$-knot.  The amplitude of each term is
essentially the product of the $A_{n}$ factors in
(\ref{H_on_s}), one for each action of $\hat H$.  Each term
can be seen as a history in which the $s$-knot state $| s'
\rangle$ evolves until it reaches $| s \rangle$.  The
resulting sum admits a graphical interpretation: The
sequence of actions of $\hat H$ on $| s' \rangle$ can be
represented by the spacetime world-sheet swept by $s'$
moving in time and undergoing a discrete transition at each
action of $\hat H$.  This world-sheet is a precisely a spin
foam.  The faces of the complex are swept out by the spin
network links and the edges by the nodes.  Faces are colored
just as the underlying link, and edges as the underlying
node.  The transitions generated by $\hat H$ given in Figure
3, is represented by the vertex illustrated in Figure 4. 
Its amplitude is given by the corresponding matrix elements
$A_{n}$ of $\hat H$, defined (\ref{H_on_s}).  For instance,
a term of order one is represented in Figure 5.  The 4d
``spacetime'' in which this evolution takes place
corresponds to the classical coordinate spacetime: a
mathematical artifact.
\begin{figure}
\centerline{\includegraphics[width=5.5cm]{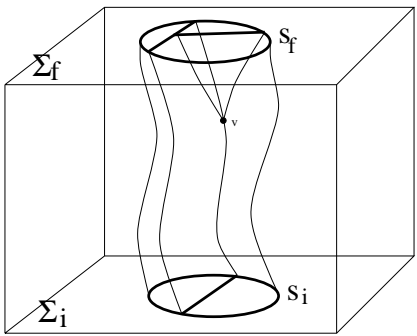}}
\centerline{
\caption[]{A first order diagram.}}
\label{order1}
\label{sei}
\end{figure}

The result is that the expansion (\ref{expansion}) of
$\langle s | P | s' \rangle$ can be written as a sum over
topologically inequivalent \emph{spin foams} $\sigma$,
bounded by the initial and final spin networks $s'$ and $s$. 
Each surface $\sigma$ represents the history of the initial
$s$-knot state, and is weighted by the product of
coefficients $A_{v}$, associated to the vertices of
$\sigma$.  That is, we obtain a spin foam model as in
(\ref{sumoverfoams}), where the vertex amplitude is
determined by the matrix elements of the hamiltonian
constraint.  

The spin foam inherits from the spin networks a geometrical
interpretation in terms of quanta of areas and volume, and can be seen
as an evolving quantum 3-geometry, that is as a quantum 4-geometry. 
In fact, a spin foam can be seen as the formalization of Weehler's
intuition of the foamy structure of Planck scale geometry
\cite{wheelerdewitt}.  In this picture, the problem of specifying the
correct hamiltonian constraint is translated into the problem of
finding the right vertex amplitude, 4d general covariance is manifest,
and we have the usual advantages of the covariant formalism.  In
particular, 4d diff-invariant quantum observables can be understood
\cite{hartle} in form more simple and intuitive than in the canonical
framework.

Before concluding, it might be instructive to compare the
spin foam formalism with the covariant formalism of a
bosonic string theory as an integral over surfaces
\begin{equation}
	Z = \int [D\sigma_{string}] \ e^{-i {\sf A}[\sigma_{string}]}\ .
	\label{eq:string}
\end{equation}
Here $\sigma_{string}$ is the embedding of the string world-sheet in
the target spacetime, and ${\sf A}[\sigma_{string}]$ is the area of
the string world-sheet, which is induced by the fixed background
metric of the target spacetime.  The spin foam partition function
(\ref{sumoverfoams}) can be easily rewritten as
\begin{equation}  
\label{sumoverfoams2} 
 Z \ \ =\ \sum_{\mbox{\scriptsize spin} \atop \mbox{\scriptsize foams}
 \, \scriptstyle \sigma} \ \ e^{-i\sum_{\mbox{\tiny vertices} 
 \atop \scriptscriptstyle{v \in \sigma}}\ \displaystyle{a_{v}}} \ .
\end{equation}
There is a similarity between (\ref{eq:string}) and
(\ref{sumoverfoams2}).  They are both sums over surface
configurations, with amplitudes which are exponentials of a local
expression on the surface.  There are however two crucial differences. 
The first is that the surfaces in (\ref{sumoverfoams2}) branch off and
carry spins, while the bosonic string world-sheets do not.  The
second, and most important difference is that the action of the
string, namely the area ${\sf A}[\sigma_{string}]$ depends on the
target space metric, and therefore depends on the precise location of
$\sigma_{string}$ in the target space.  The sum over surfaces in
(\ref{eq:string}) is thus an integral in which two slightly deformed
locations of $\sigma_{string}$ in the target space count as different. 
On the contrary, in (\ref{sumoverfoams2}) no manifold background
structure enters the action (the exponent), and the weight does not
depend on the location of the spin foam in the manifold, but only on
its diff-invariant features.  Accordingly the sum over surfaces is
over diff inequivalent classes only, and it is a sum and not an
integral.

It has been suggested, on the other hand, that the background
independent, general relativistic, formulation of string theory, whose
lack is the most serious open problem in that theory, could look like
a spin foam model \cite{leestring}.  For instance, (\ref{eq:string})
could perhaps be obtained as an approximation of an expression of the
form (\ref{sumoverfoams2}), when expanding small fluctuations
$\sigma_{fluctuation}$ of $\sigma$ over a ``background''
$\sigma_{background}$, representing the background geometry.  The area
in (\ref{eq:string}) could then emerge as the result of local diff
invariant interactions between $\sigma_{fluctuation}$ and
$\sigma_{background}$.  Since $\sigma_{background}$ can be interpreted
as a 4-geometry and the area is the lowest dimension additive
functional for a surface, it is not unlikely that an expression like
(\ref{eq:string}) could emerge from (\ref{sumoverfoams2}).

\bigskip

In conclusion, the picture of the theory as a sum over spin foams is
thus common to both loop quantum gravity and the covariant approaches:
a common and compelling formalism for general relativistic quantum
field theory seems to be emerging from different quarters.  It is
remarkably different from the formalism of non general relativistic
quantum field theory.  The spin foam formalism is explicitly diff
invariant and background independent.  Diff invariance is strictly
related to the short scale discreteness of the theory, which is
reflected in the formalism by the purely combinatorial nature of
(\ref{sumoverfoams2}).  In turn, this fundamental short scale
discreteness seems to be capable of taking care of the conventional
quantum field theory ultraviolet divergences.

\section{Conclusion} 

Hopefully, the picture that I have described and which is emerging
from different quarters, some variant of this picture, or maybe a
different picture, will lead us to the definition of a consistent
quantum general relativistic theory with general relativity as its
classical limit, and thus to a successful conclusion of the twentieth
century revolution in fundamental physics.  

Such a successful conclusion, in my opinion, is not to be searched in
a omni-comprehensive Lagrangian, or in a theory of everything. 
Rather, it is to be searched in a new conceptual framework for
describing the physical world at the fundamental level.  A conceptual
framework capable of replacing the extraordinary powerful Newtonian
framework, but taking into account the deep modification of the basic
notions of matter, causality, space and time that represent a major
legacy of this tormented century. 

The ``great'' scientific revolution, the seventeenth's century one,
was opened by Copernicus' {\em De Revolutionibus\/} and successfully
concluded by Newton's {\em Principia}.  That is to say, it lasted a
century and a half.  The current revolution has still a chance to be a
bit shorter.


\begin{thebibliography}{99}

\bibitem{RovelliSmolin95} C Rovelli and L Smolin,
``Discreteness of Area and Volume in Quantum Gravity'', Nucl
Phys B442 (1995) 593, gr-qc/9411005; Erratum: Nucl Phys B456
(1995) 734.

\bibitem{RovelliSmolin95b} C Rovelli and L Smolin, ``Spin
Networks and Quantum Gravity'', Phys Rev D52 (1995) 5743,
gr-qc/9505006.

\bibitem{Atiyah} MF Atiyah, ``Topological quantum field
theories'', Publ Math Inst Hautes Etudes Sci, Paris 68
(1989) 175; {\em The Geometry and Physics of Knots},
Accademia Nazionale dei Lincei, (Cambridge University Press,
1990).  E Witten, Comm Math Phys 117 (1988) 353.

\bibitem{Turaev-Viro} VG Turaev and OY Viro, ``State sum
invariants of 3-manifolds and quantum $6j$ symbols'',
Topology 31 (1992) 865.  VG Turaev, ``Quantum Invariants of
Knots and 3-manifolds'' (de Gruyter, New York 1994).

\bibitem{Turaev} V Turaev, ``Quantum invariants of
3-manifolds and a glimpse of shadow topology'' in {\em
Quantum Groups}, Springer Lecture Notes in Mathematics 1510,
pp 363-366 (Springer-Verlag, New York, 1992); {\em Quantum
Invariants of Knots and 3-Manifolds} (de Gruyter, New York,
1994).

\bibitem{Ooguri:1992b} H Ooguri, ``Topological lattice
models in four dimensions'', Mod Phys Lett A7 (1992) 2799.

\bibitem{Crane-Yetter} L Crane and D Yetter, ``A Categorical
construction of 4-D topological quantum field theories'', in
{\it Quantum Topology}, L Kaufmann and R Baadhio eds (World
Scientific, Singapore 1993), hep-the/9301062.

\bibitem{2d} F David, Nucl Phys B257 (1985) 45.  J Ambjorn,
B Durhuus, J Fr\"olich, Nucl Phys B257 (1985) 433.  VA
Kazakov, IK Kostov, AA Migdal, Phys Lett 157 (1985) 295.  DV
Boulatov, VA Kazakov, IK Kostov, AA Migdal, Nucl Phys B275
(1986) 641.  M Douglas, S Shenker, Nucl Phys B335 (1990)
635.  D Gross, A Migdal, Phys Rev Lett 64 (1990) 635;
E Brezin, VA Kazakov, Phys Lett { B236} (1990) 144.
   
\bibitem{Boulatov:1992} 
DV Boulatov, Mod Phys Lett A7 (1992) 1629.
  
\bibitem{Baez}
J Baez, ``Spin foam models'', Class Quant Grav 15 (1998)
1827.

\bibitem{Reisenberger97} M P Reisenberger, ``A lattice
worldsheet sum for 4-d Euclidean general relativity'',
gr-qc/9711052.

\bibitem{Barrett:1998} J W Barrett and L Crane,
``Relativistic spin networks and quantum gravity'', J Math
Phys 39 (1998) 3296.
 
\bibitem{othersn} F Markopoulou, L Smolin Phys Rev
D58:084032 (1998).  L Smolin, hept-the/9001022,
hep-the/9808192.  R De Pietri, ``Canonical `loop' quantum
gravity and spin foam models'', to appear in the proceedings
of the XXIIIth Congress of the Italian Society for General
Relativity and Gravitational Physics (SIGRAV), 1998,
gr-qc/9903076.  J W Barrett and L Crane, ``A Lorentzian
signature model for quantum general relativity'',
gr-qc/9904025.  J Iwasaki, ``A surface theoretic model for
quantum gravity'', gr-qc/9903122.

\bibitem{ReisenbergerRovelli97}
C Rovelli: ``Quantum gravity as a `sum over surfaces' `',
Nucl Phys B (Proc Suppl) 57 (1997) 28.  M Reisenberger and C
Rovelli, ``\ `Sum over Surfaces' form of Loop Quantum
Gravity'', Phys Rev D56 (1997) 3490, gr-qc/9612035.  C
Rovelli, ``The projector on physical states in loop quantum
gravity'', Phys Rev D59 (1999) 104015, gr-qc/9806121.

\bibitem{Rovelli97b} C Rovelli, ``Strings, loops and others:
A Critical survey of the present approaches to quantum
gravity'', in {\em Gravitation and Relativity at the turn
of the millennium.  Proceedings of the GR-15 conference,
IUCAA, Pune, India December 1997} (Inter University center
for Astronomy and Astrophysics Pune 1998), gr-qc/9803024.

\bibitem{Rovelli91b} C Rovelli, ``What is observable in
classical and quantum gravity?''  Class Quant Grav 8 (1991)
297.

\bibitem{Rovelli97c} C Rovelli, ``Halfway Through the Woods:
Contemporary Research on Space and Time'', in {\it The
Cosmos of Science}, J Earman and J D Norton eds (University
of Pittsburgh Press and Universit\"ats-Verlag Konstanz,
Pittsburgh and Konstanz 1997).

\bibitem{Rovelli99} C Rovelli, ``Quantum spacetime: what do
we know?'', in {\em Physics meets philosophy at the Planck
scale}, C Callender N Hugget eds (Cambridge University
Press, in print) gr-qc/9903045.


\bibitem{time} C Rovelli, ``Time in quantum gravity: an
hypothesis'', Phys Rev D43 (1991) 442; ``Quantum mechanics
without time: a model'', Phys Rev D42 (1990) 2638.

\bibitem{termo} C Rovelli: ``Statistical mechanics of
gravity and thermodynamical origin of time", Class and
Quantum Grav 10 (1993) 1549; ``The statistical state of the
universe", Class and Quantum Grav 10 (1993) 1567.

\bibitem{alain} A Connes and C Rovelli: ``Von Neumann
algebra automorphisms and time versus thermodynamics
relation in general covariant quantum theories'', Class and
Quantum Grav 11 (1994) 2899.

\bibitem{qm} C Rovelli: ``Relational Quantum Mechanics", Int
J Theor Phys 35 (1996) 1637.

\bibitem{Haag} R Haag, D Kastler ``An algebraic approach to
quantum field theory'', J Math Phys 5 (1964) 848.  R Haag,
{\em Local Quantum Physics} (Springer Verlag, Berlin 1992).

\bibitem{hartle} J B Hartle, ``Spacetime Quantum Mechanics
and the Quantum Mechanics of Spacetime'' in {\sl Gravitation
and Quantizations}, Proceedings of the 1992 Les Houches
Summer School, ed by B Julia and J Zinn-Justin, Les Houches
Summer School Proceedings, Vol LVII (North Holland,
Amsterdam, 1995); gr-qc/9304006.

\bibitem{Ish94} C J Isham, ``Quantum Logic and the Histories
Approach to Quantum Theory'', J Math Phys 35 (1994) 2157,
gr-qc/9308006.  CJ Isham, J Butterfield, ``Some Possible
Roles for Topos Theory in Quantum Theory and Quantum
Gravity'' {\em Foundations of Physics}, to appear,
gr-qc/9910005.

\bibitem{tomita} M Tomita, V-th Functional Analysis
Symposium of the Math Soc of Japan Sendai (1967).  M
Takesaki {\em Tomita's Theory of Modular Hilbert Algebras
and its Applications}, Springer Lecture Notes 128
(Springer-Verlag, Berlin).

\bibitem{connes} A Connes, ``Une classification des facteurs
de type III'', Ann Sci Ecole Norm Sup 6 (1973) 133.

\bibitem{wheelerdewitt} J Wheeler, in {\em Battelle Rencontres
1967}, C DeWitt and J A Wheeler eds (Benjamin, New York, 
1968).  B DeWitt, Phys Rev D160, 1113 (1967).

\bibitem{hawk} SW Hawking, ``The Path-Integral Approach to
Quantum Gravity'', in {\sl General Relativity: An Einstein
Centenary Survey}, SW Hawking and W Israel eds (Cambridge
University Press, Cambridge 1979).

\bibitem{hh} JB Hartle, SW Hawking, Phys Rev D28 (1983) 2960. 

\bibitem{oldloop} S Mandelstam, Ann Phys 19 (1962) 1.  AM
Polyakov, Phys Lett 82B (1979) 247; Nucl Phys B164 (1979)
171.  YM Makeenko, AA Migdal Phys Lett 88B (1979) 135.  Y
Nambu, Phys Lett 80B (1979) 372; F Gliozzi, T Regge, MA
Virasoro, Phys Lett 81B (1979) 178; M Virasoro, Phys Lett
82B (1979) 436; A Jevicki, B Sakita, Phys Rev D22 (1980)
467; B Sakita, ``Collective field theory'' in {\em Field
theory in elementary particles}, B Kursunoglu and A
Perlmutter eds (Plenum Publishing Corporation, 1983).  K
Wilson, Phys Rev D10 (1974) 247; J Kogut, L Suskind, Phys
Rev D11 (1975) 395; L Susskind ``Coarse grained quantum
chromodynamics'' in {\it Les Houches XXIX Weak and
Electromagnetic Interactions at High Energy}, R Balian and C
H Llewellyn-Smith eds (North Holland, Amsterdam).

\bibitem{loop} C Rovelli, ``Loop quantum gravity'', in {\em
Living Reviews in Relativity\/}, electronic journal,
http://www.livingreviews.org, gr-qc/9710008.

\bibitem{Karpacz} C Rovelli, M Gaul, to appear in the proceedings
of the {\em XXXV Karpacz Winter School on Theoretical Physics:
From Cosmology to Quantum Gravity}, Polanica, Poland, 1999. 
gr-qc/xxxxxxx. 

\bibitem{Abhaybook} A Ashtekar, {\em Lectures on non-perturbative
canonical gravity\/} (World Scientific, Singapore 1991).

\bibitem{Rovelli91} C Rovelli, ``Ashtekar formulation of general
relativity and loop space nonperturbative quantum gravity: A Report'',
Class Quant Grav 8 (1991) 1613.

\bibitem{Barbero} F Barbero, ``Real Ashtekar Variables for
Lorentzian Signature Space-times'' Phys Rev D51 (1995) 5507,
gr-qc/9410014; ``From Euclidean to Lorentzian General
Relativity: The Real Way'', Phys Rev D54 (1996) 1492,
gr-qc/9605066.

\bibitem{Ashtekar93} A Ashtekar, CJ Isham, ``Representations
of the holonomy algebra of gravity and non-abelian gauge
theories'', Class and Quantum Grav 9 (1992) 1069.  A
Ashtekar, ``Mathematical Problems of Non-perturbative
Quantum General Relativity'', in: J Zinn-Justin and B Julia
(eds ), ``Les Houches Summer School on Gravitation and
Quantizations,'' Les Houches, France, 1992 (North-Holland,
Amsterdam 1995), gr-qc/9302024.

\bibitem{Lewand_Karpacz99} J Lewandowski, to appear on the
proceedings of the XXXV Karpacz Winter School on Theoretical
Physics: ``From Cosmology to Quantum Gravity,'' Polanica,
Poland, February 2--12, 1999.

\bibitem{areaE} C Rovelli: ``Area is the length of
Ashtekar's triad field", Phys Rev D47 (1993) 1703.

\bibitem{area} C Rovelli: ``A generally covariant quantum
field theory and a prediction on quantum measurements of
geometry", Nucl Phys B405 (1993) 797.

\bibitem{leearea} L Smolin, ``Finite, diffeomorphism
invariant observables in quantum gravity'' Phys Rev D49
(1994) 4028.

\bibitem{Smolin97} L Smolin, ``The future of spin
networks'', gr-qc/9702030.

\bibitem{DePietriRovelli96} R De Pietri and C Rovelli,
``Geometry Eigenvalues and Scalar Product from Recoupling
Theory in Loop Quantum Gravity'', Phys Rev D54 (1996) 2664,
gr-qc/9602023.

\bibitem{moduli} N Grott and C Rovelli: ``Moduli spaces
structure of knots with intersections", J Math Phys 37
(1996) 3014.

\bibitem{zap} J Zapata, J Math Phys 38 (1997) 5663, Gen Rel
Grav 30 (1998) 1229.

\bibitem{Thiemann96} T Thiemann, ``Anomaly-free formulation
of non-perturbative, four-dimensional Lorentzian quantum
gravity'', Phys Lett B380 (1996) 257; ``Quantum Spin
Dynamics (QSD)'', Class Quant Grav 15 (1998) 839,
gr-qc/9606089.

\bibitem{RovelliSmolin90} C Rovelli and L Smolin, ``Loop
representation of quantum general relativity'', Nucl Phys
B331 (1990) 80.

\bibitem{outline} C Rovelli, ``Outline of a general
covariant quantum field theory and a quantum theory of
gravity'', J Math Phys 36 (1996) 6529.

\bibitem{Borissov97} R Borissov, R DePietri, C Rovelli:
``Matrix elements of Thiemann hamiltonian'', Classical and
Quantum Gravity 14 (1997) 2793; gr-qc/97703090

\bibitem{RovelliSmolin94} C Rovelli and L Smolin, ``The
physical hamiltonian in nonperturbative quantum gravity'',
Phys Rev Lett 72 (1994) 446, gr-qc/9308002.

\bibitem{blackhole} C Rovelli: ``Black Hole Entropy from
Loop Quantum Gravity", Phys Rev Lett 14 (1996) 3288; ``Loop
Quantum Gravity and Black Hole Physics", Helvetica Physica
Acta, 69 (1996) 582; gr-qc/9608032.  K Krasnov,
``Geometrical entropy from loop quantum gravity'' Phys Rev
D55 (1997) 3505.  A Ashtekar, J Baez, A Corichi, K Krasnov,
``Quantum Geometry and Black Hole Entropy'', Phys Rev Lett
80 (1998) 904-907.

\bibitem{witten}
E Witten ``(2+1) dimensional gravity as an exactly soluble system''. 
Nucl Phys B311 (1988) 46.

\bibitem{carlip} S Carlip, {\em Quantum Gravity in 2+1
Dimensions} (Cambridge University Press, Cambridge, 1998).

\bibitem{3dnoi} A Ashtekar, V Husain, J Samuel, C Rovelli, L
Smolin, ``2+1 quantum gravity as a toy model for the 3+1
theory'', Class and Quantum Grav 6 (1989) L185.

\bibitem{kl} L Kauffman, S Lins, {\em Temperley-Lieb
Recoupling Theory and Invariants of 3-Manifolds}, (Princeton
University Press, Princeton 1994).

\bibitem{crane} L Crane, D Yetter, ``On algebraic structure
implicit in topological quantum field theory'', Kansas
preprint, 1994.  D Yetter ``State sum invariants of
3-manifolds associated to Artinian tortile categories'',
Topology and its Applications 58 (1994), 47.

\bibitem{fotini} F Markopoulou ``Dual formulation of spin
network evolution'', gr-qc/9704013.

\bibitem{Ponzano:1968} G Ponzano and T Regge, in {\em
Spectroscopy and group theoretical methods in Physics},
F Bloch ed (North-Holland, Amsterdam, 1968).

\bibitem{BF} G Horowitz, ``Exactly soluble diffeomorphism
invariant theories'', Comm Math Phys 125 (1989) 417.  M Blau
and G Thompson, ``Topological gauge theories of
antisymmetric tensor fields'', Phys Lett B228 (1989) 64.  J
Baez, ``Four dimensional $BF$ theory as a topological
quantum field theory'', Lett Math Phys 38 (1996) 129.  A
Cattaneo, P Cotta-Ramusino, J Fr\"ohlich and M Martellini
``Topological $BF$ theories in 3 and 4 dimensions'', J Math
Phys 36 (1995) 6137.

\bibitem{oogurirovelli} C Rovelli: ``Basis of the
Ponzano-Regge-Turaev-Viro-Ooguri quantum gravity model is
the loop representation basis'', Phys Rev D48 (1993) 2702.

\bibitem{foxon} TJ Foxon, ``Spin networks, Turaev-Viro
theory and the loop representation'', Class Quant Grav 12
(1995) 951-964, gr-qc/9408013.

\bibitem{Crane} L Crane, L Kauffman and D Yetter, ``State
sum invariants of 4-manifolds'', J Knot Theory and
Ramifications 6 (1997) 177, hep-th/9409167.
 
\bibitem{DF} R De Pietri and L Freidel, Class Quant Grav 16
(1999) 2187, gr-qc/9804071.  M Reisenberger, Class Quant
Grav 16 (1999) 1357, gr-qc/9804061.
  
\bibitem{Barbieri} A Barbieri, Nucl Phys B518 (1998) 714.  L
Freidel and K Krasnov, Class Quantum Grav 16 (1999) 351.  J
Baez and J Barrett, ``The Quantum Tetrahedron in 3 and 4
Dimensions'', gr-qc/9903060.

\bibitem{altridue} J W Barrett and R Williams, ``The
asymptotic of an amplitude for the 4-simplex'',
gr-qc/9809032.
	
\bibitem{FW} L Freidel and K Krasnov, ``Spin Foam Models and
the Classical Action Principle'', Adv Theor Math Phys 2
(1998) 1221.

\bibitem{DKFR} R De Pietri, L Freidel, K Krasnov, C Rovelli,
``Barrett-Crane model from a Boulatov-Ooguri field theory
over a homogeneous space'',
 hep-the/9907154
		  
\bibitem{Baez:1994} J Baez, ``Strings, loops, knots and
gauge fields'', in {\em Knots and Quantum Gravity}, J
Baez ed (Oxford University Press, Oxford 1994).

\bibitem{Reisenberger} M Reisenberger, ``Worldsheet
formulations of gauge theories and gravity'', talk given at
the 7th Marcel Grossmann Meeting Stanford, July 1994,
gr-qc/9412035.
	
\bibitem{Iwasaki} J Iwasaki, ``A definition of the
Ponzano-Regge quantum gravity model in terms of surfaces'',
Journ Math Phys 36 (1995) 6288.
 
\bibitem{Baez2} J Baez, ``An introduction to spin foam
models of BF theory and quantum gravity'', to appear in to
appear in {\em Geometry and Quantum Physics}, H Gausterer
and H Grosse eds, Lecture Notes in Physics (Springer-Verlag,
Berlin), gr-qc/9905087. 

\bibitem{Thiemanint} T Thiemann, ``Kinematical Hilbert
Spaces for Fermionic and Higgs Quantum Field Theories'',
Class Quantum Grav 15 (1998) 1487, gr-qc/9705021.

\bibitem{Baez98} J C Baez, ``Spin Foam Models'', Class Quant
Grav 15 (1998) 1827, gr-qc/9709052.

\bibitem{leestring}
L Smolin, ``Towards a background independent approach to M theory'',
hep-the/9808192; ``Strings as perturbations of evolving
spin-networks'', hep-the/9801022.
		  
\end{thebibliography}
\end{document}